\documentclass{vldb}
\usepackage{graphicx}
\usepackage{booktabs} 
\usepackage[colorinlistoftodos]{todonotes}
\usepackage{algorithmic}
\usepackage[linesnumbered, ruled, vlined]{algorithm2e}
\usepackage{bm}
\usepackage{booktabs}
\usepackage{mathrsfs}
\usepackage{multirow}

\usepackage[utf8]{inputenc}
\usepackage[T1]{fontenc}
\usepackage{hyperref} 
\usepackage{url}     
\usepackage{amsfonts}       
\usepackage{nicefrac}       
\usepackage{microtype}
\usepackage{graphicx}
\usepackage{amsmath}
\usepackage{epsfig}
\usepackage{amsfonts}
\usepackage[misc]{ifsym}
\usepackage{bbding}
\usepackage{colortbl}
\usepackage{enumitem}
\usepackage{times}

\newtheorem{theorem}{Theorem}

\newcommand{\bz}{{\mathbf z}}
\newcommand{\bh}{{\mathbf h}}
\newcommand{\bff}{{\mathbf f}}
\newcommand{\bdelta}{{\mathbf \delta}}
\newcommand{\bsigma}{{\mathbf \sigma}}
\newcommand{\bmu}{{\mathbf \mu}}

\newcommand{\G}{\mathcal{G}}
\newcommand{\V}{\mathcal{V}}
\newcommand{\E}{\mathcal{E}}
\newcommand{\W}{\mathcal{W}}
\newcommand{\T}{\mathcal{T}}
\newcommand{\F}{\mathcal{F}}
\newcommand{\A}{\mathcal{A}}
\newcommand{\BH}{\mathbf{H}}
\newcommand{\BX}{\mathbf{X}}
\newcommand{\BZ}{\mathbf{Z}}
\newcommand{\BS}{\mathbf{S}}
\newcommand{\bx}{\mathbf{x}}
\newcommand{\by}{\mathbf{w}}
\newcommand{\bv}{\mathbf{h}_v}

\newcommand{\AliG}{\textit{AliGraph} }

\newcommand{\tabincell}[2]{\begin{tabular}{@{}#1@{}}#2\end{tabular}}  

\definecolor{mygray}{gray}{.9}
\definecolor{myred}{rgb}{0.8902,0.09,0.051}
\definecolor{mygreen}{rgb}{0.188, 0.502, 0.078}
\definecolor{myemph}{rgb}{1,0.890, 0.518}

\newcommand{\RSymbol}{\color{mygreen}{{\CheckmarkBold}}}
\newcommand{\WSymbol}{\color{myred}{\scriptsize{\XSolidBold}}}

\vldbTitle{AliGraph}
\vldbDOI{https://doi.org/TBD}

\begin{document}

\title{{\ttlit AliGraph}:
A Comprehensive Graph Neural Network Platform}

\author{
\alignauthor{
    Rong Zhu, Kun Zhao, Hongxia Yang, Wei Lin, Chang Zhou, Baole Ai, Yong Li, Jingren Zhou \\
}
\affaddr{
    \medskip
    \large{
   \bf{Alibaba Group}}
   \vspace{-3em}
}
}

\maketitle

\begin{abstract}
An increasing number of machine learning tasks require dealing with large graph datasets, which capture rich and complex relationship among potentially billions of elements. Graph Neural Network (GNN) becomes an effective way to address the graph learning problem by converting the graph data into a low dimensional space while keeping both the structural and property information to the maximum extent and constructing a neural network for training and referencing. However, it is challenging to provide an efficient graph storage and computation capabilities to facilitate GNN training and enable development of new GNN algorithms. In this paper, we present a comprehensive graph neural network system, namely {\textit{AliGraph}}, which consists of distributed graph storage, optimized sampling operators and runtime to efficiently support not only existing popular GNNs but also a series of in-house developed ones for different scenarios. The system is currently deployed at Alibaba to support a variety of business scenarios, including product recommendation and personalized search at Alibaba's E-Commerce platform. By conducting extensive experiments on a real-world dataset with 492.90 million vertices, 6.82 billion edges and rich attributes, \AliG performs an order of magnitude faster in terms of graph building (5 minutes vs hours reported from the state-of-the-art PowerGraph platform). At training, \AliG runs 40\%-50\% faster with the novel caching strategy and demonstrates around 12 times speed up with the improved runtime. In addition, our in-house developed GNN models all showcase their statistically significant superiorities in terms of both effectiveness and efficiency (e.g., 4.12\%--17.19\% lift by F1 scores).
\end{abstract}

\section{Introduction} \label{sec:intro}

As a sophisticated model, graph has been widely used to model and manage data in a wide variety of real-world applications. Typical examples include social networks~\cite{Hamilton2017Representation,kipf2017semi},
physical systems~\cite{Battaglia:2016,Sanche:2018},
biological networks~\cite{Fout:2017}, knowledge graphs \cite{Hamaguchi:2017} and etc~\cite{Elias:2017}. Graph analytics, which explore underlying insights hidden in graph data, have drawn significant research attention in the last decade. They have been witnessed to play important roles in numerous areas, i.e., node classification~\cite{Bhagat2011Node}, link prediction~\cite{Liben2003The}, graph clustering~\cite{Brandes2003Experiments}, recommendation~\cite{Wang2010Graph}, among many others.

As conventional graph analytic tasks often suffer from high computation and space costs~\cite{cui2018survey, Hamilton2017Representation}, a new paradigm, called \emph{graph embedding} (GE), paves an efficient yet effective way to address such problems. Specifically, GE converts the graph data into a low-dimensional space such that the structural and content information in the graph can be preserved to the maximum extent. After that, the generated embeddings are fed as features into the downstream machine learning tasks. Furthermore, by incorporating with deep learning techniques, \emph{graph neural networks} (GNN) are proposed by integrating GE with \emph{convolutional neural network} (CNN)~\cite{kipf2017semi,chen2018fastgcn,huang2017accelerated,GraphSage:HamiltonYL17}. In CNN, shared weights and multi-layered structure are applied to enhance its learning power \cite{LeCun:2015}. And graphs are the most typical locally connected structures, with shared weights to reduce the computational cost and the multi-layer structure being the key to deal with hierarchical patterns while capturing features of various sizes. GNNs find such generalizations of CNNs to graphs. Thus, GNN not only embraces the flexibility of GE but also showcases its superiority in terms of both effectiveness and robustness with generalizations of CNNs.

\smallskip
\noindent{\textbf{\underline{Challenges.}}}
In the literature, considerable research efforts have been devoted in developing GE and GNN algorithms. These works mainly concentrate on simple graphs with no or little auxiliary information. However, the rising of big data and complex systems reveal new insights in graph data. As a consensus~\cite{cui2018survey, Hamilton2017Representation, Cai2017A, Goyal2018Graph}, the vast majority of graph data related to real-world commercial scenarios exhibits four properties, namely \emph{large-scale}, \emph{heterogeneous}, \emph{attributed} and \emph{dynamic}. For example, nowadays e-commerce graphs often contain billions of vertices and edges with various types and rich attributes, and quickly evolve over time. These properties bring great challenges for embedding and representing graph data as follows:

\begin{itemize}[leftmargin=*]
	\item 
	The core steps in GNN are particularly optimized for grid structures, such as images, but not feasible for graphs in irregular Euclidean space. Thus, existing GNN methods can not scale on real-world graphs with exceedingly large sizes.
	\textbf{\textit{The first problem is how to improve the time and space efficiencies of GNN on large-scale graphs}}?
	
	\item
	Different types of objects characterize the data from multiple perspectives. They provide richer information but increase the difficulty to map the graph information into a \emph{singleton} space. Thus, \textbf{\textit{the second problem is how to elegantly integrate the heterogeneous information to be an unified embedding result}}? 
	\item
	The attribute information can further enhance the 
	power of the embedding results and make inductive GE possible~\cite{cui2018survey, Hamilton2017Representation, Cai2017A, Goyal2018Graph}. Without considering attribute information, the algorithms can only consider the transductive settings and ignore the need for predicting unseen instances.
	However, the topological structure information and unstructured
	attribute information are usually presented in two different spaces. Thus, \textbf{\textit{the third problem is how to unify them to define the information to be preserved}}?
	
	\item
	As GNN suffers from low efficiency, recomputing the embedding results from scratch with respect to structural and contextual updates are expensive. Thus, \textbf{
		\textit{the fourth problem is how to design efficient incremental GNN methods on dynamic graphs}}?
\end{itemize}

\smallskip
\noindent{\textbf{\underline{Contributions.}}}
To tackle with the above challenges, considerable research efforts have been devoted to design efficient yet effective GNN methods. In Table~\ref{tab:methods}, we categorize a series of popular GE and GNN models according to the aspects that they focus on, as well as our in-house developed models shaded in yellow. As shown, most existing methods concentrate on one or two properties at the same time. However, real-world commercial data usually faces more challenges. To mitigate this situation, in this paper, we present a comprehensive and systemic solution to GNN. We design and implement a platform, called \textit{AliGraph}, which provides both system 
and algorithms to tackle more practical problems that exhibit the four summarized arising problems, to well support a variety of GNN methods and applications. The main contributions are summarized as follows. 

\smallskip
\noindent{\textit{\textbf{System.}}}
In the underlying components of \textit{AliGraph}, we build a system to support GNN algorithms and applications. The system architecture is abstracted from general GNN methods, which consists of a \textit{storage} layer, a \textit{sampling} layer and an \textit{operator} layer. Specifically, the storage layer applies three novel techniques, namely structural and attributed specific storage, graph partition and caching neighbors of some important vertices, to store large-scale raw data to fulfill the fast data access requirements of high-level operations and algorithms. The sampling layer optimizes the key sampling operation in GNN methods.
We categorize sampling methods into three classes, namely \textsc{traverse}, \textsc{neighborhood}  and \textsc{negative} sampling, and propose lock-free methods to perform sampling operations in distributed environment.
The operator layer provides optimized implementation of two common applied operators in GNN algorithms, namely \textsc{aggregate} and \textsc{combine}. We apply a cache strategy to store some intermediate results to accelerate the computation process. These components are co-designed and co-optimized to make the whole system effective and scalable.

\smallskip
\noindent{\textit{\textbf{Algorithms.}}}
The system provides a flexible interface to design GNN algorithms. We show that all existing GNN methods can be easily implemented upon our system.
Besides, we also in-house developed several new GNNs for practical requirement and detail six here. As illustrated in Table~\ref{tab:methods}, our in-house developed methods are shaded in yellow and each of them are more flexible and practical to deal with  real-world problems. 

\begin{figure}[ht]
	\centering
	\includegraphics[width=3.1in]{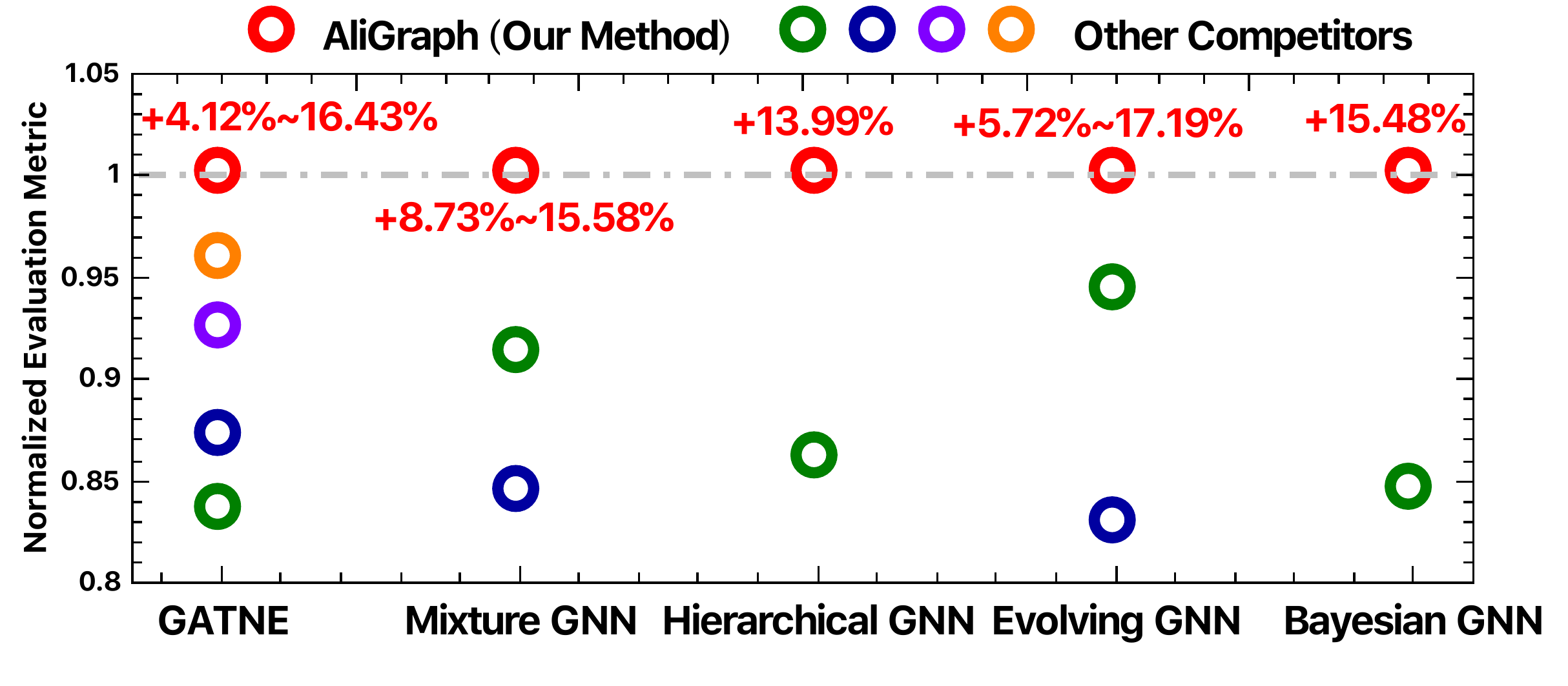}
	\vspace{-1em}
	\caption{Comparison of normalized evaluation metric on the effectiveness of different methods. The red text indicates the lifting range of each in-house developed method w.r.t.~its competitors.
	}
	\label{fig:NEM}
	\vspace{-2em}
\end{figure}

\smallskip
\noindent{\textit{\textbf{Evaluation.}}}
Our AliGraph platform is practically deployed in the Alibaba company. The experimental results verify  its effectiveness and efficiency from both the system and algorithm aspects. As shown in Figure~\ref{fig:NEM}, our  in-house developed GNN models on the AliGraph platform improves the normalized evaluation metrics by $4.12\%$--$17.19\%$. The data are collected from Alibaba's e-commerce platform   Taobao  and we contribute this dataset to the community to nourish further development\footnote{\url{https://tianchi.aliyun.com/dataset/dataDetail?dataId=9716.}}.

\begin{table}[t]
	\small
	\centering
	\caption{\label{tab:methods} The property of different methods.}
	\resizebox{\columnwidth}{!}{
		\begin{tabular}{c|c|c|c|c|c|c}
			\hline
			\rowcolor{mygray}
			&  & \multicolumn{2}{c|}{Heterogeneous}
			&  &  &  \\
			\cline{3-4}
			\rowcolor{mygray}
			\multirow{-2}{*}{Category} &\multirow{-2}{*}{Method} & Node & Edge & \multirow{-2}{*}{Attributed} & \multirow{-2}{*}{Dynamic} & \multirow{-2}{*}{Large-Scale}\\
			\hline
			\rowcolor{white}
			\multirow{19}{*}{\tabincell{c}{Classic \\ Graph \\ Embedding}} & DeepWalk & \WSymbol  & \WSymbol & \WSymbol & \WSymbol & \WSymbol\\
			& Node2Vec & \WSymbol  & \WSymbol & \WSymbol & \WSymbol & \WSymbol\\
			& LINE & \WSymbol & \WSymbol & \WSymbol & \WSymbol & \WSymbol \\
			& NetMF & \WSymbol & \WSymbol & \WSymbol & \WSymbol & \WSymbol \\
			& TADW & \WSymbol & \WSymbol & \RSymbol & \WSymbol & \WSymbol \\
			& LANE & \WSymbol  &  \WSymbol & \RSymbol & \WSymbol & \WSymbol \\
			& ASNE & \WSymbol  & \WSymbol  & \RSymbol & \WSymbol & \WSymbol\\
			& DANE & \WSymbol  &  \WSymbol & \RSymbol & \WSymbol & \WSymbol\\
			& ANRL & \WSymbol  &  \WSymbol & \RSymbol & \WSymbol & \WSymbol\\
			& PTE & \WSymbol  & \RSymbol & \WSymbol & \WSymbol & \WSymbol \\
			& Methpath2Vec & \WSymbol  & \RSymbol & \WSymbol  &  \WSymbol & \WSymbol \\
			& HERec & \WSymbol & \RSymbol &  \WSymbol & \WSymbol & \WSymbol \\
			& HNE & \WSymbol  & \WSymbol  &  \WSymbol & \WSymbol & \WSymbol  \\
			& PMNE & \WSymbol  & \RSymbol & \RSymbol &  \WSymbol & \WSymbol \\
			& MVE & \WSymbol  & \RSymbol & \RSymbol &  \WSymbol & \WSymbol \\
			& MNE & \WSymbol & \RSymbol & \RSymbol & \WSymbol & \WSymbol  \\
			& Mvn2Vec & \WSymbol  & \RSymbol & \RSymbol & \WSymbol & \WSymbol  \\
			\hline
			\hline
			\cellcolor{white}
			\multirow{11}{*}{GNN} & \cellcolor{white} Structural2Vec & \cellcolor{white} \WSymbol &  \cellcolor{white} \WSymbol & \cellcolor{white} \RSymbol & \cellcolor{white} \WSymbol & \cellcolor{white} \WSymbol \\
			\cellcolor{white}
			& GCN & \cellcolor{white} \WSymbol  &  \cellcolor{white} \WSymbol & \cellcolor{white} \RSymbol &   \cellcolor{white} \WSymbol & \cellcolor{white} \WSymbol\\
			& FastGCN & \cellcolor{white} \WSymbol  &  \cellcolor{white} \WSymbol & \cellcolor{white} \RSymbol &   \cellcolor{white} \WSymbol & \cellcolor{white} \WSymbol\\
			& AS-GCN & \cellcolor{white} \WSymbol  &  \cellcolor{white} \WSymbol & \cellcolor{white} \RSymbol &   \cellcolor{white} \WSymbol & \cellcolor{white} \WSymbol\\
			& GraphSAGE & \cellcolor{white} \WSymbol  &  \cellcolor{white} \WSymbol & \cellcolor{white} \RSymbol &   \cellcolor{white} \WSymbol & \cellcolor{white} \WSymbol\\
			& \cellcolor{myemph}\textsf{HEP} & \cellcolor{myemph}\RSymbol  & \cellcolor{myemph}\RSymbol & \cellcolor{myemph}\RSymbol & \cellcolor{myemph}\WSymbol & \cellcolor{myemph}\WSymbol  \\
			& \cellcolor{myemph}\textsf{AHEP} & \cellcolor{myemph}\RSymbol  & \cellcolor{myemph}\RSymbol & \cellcolor{myemph}\RSymbol & \cellcolor{myemph}\WSymbol & \cellcolor{myemph}\RSymbol  \\
			& \cellcolor{myemph} \textsf{GATNE} & \cellcolor{myemph}\RSymbol & \cellcolor{myemph}\RSymbol & \cellcolor{myemph}\RSymbol &  \cellcolor{myemph}\WSymbol & \cellcolor{myemph}\RSymbol \\
			& \cellcolor{myemph} \textbf{Mixture GNN} & \cellcolor{myemph} \RSymbol & \cellcolor{myemph}\RSymbol & \cellcolor{myemph}\RSymbol &
			\cellcolor{myemph} \WSymbol & \cellcolor{myemph} \WSymbol \\
			& \cellcolor{myemph} \textbf{Hierarchical GNN} & \cellcolor{myemph} \RSymbol & \cellcolor{myemph} \RSymbol & \cellcolor{myemph} \RSymbol & \cellcolor{myemph} \WSymbol & \cellcolor{myemph} \WSymbol \\
			& \cellcolor{myemph} \textbf{Bayesian GNN} & \cellcolor{myemph} \WSymbol  & \cellcolor{myemph} \RSymbol & \cellcolor{myemph} \RSymbol & \cellcolor{myemph} \WSymbol & \cellcolor{myemph} \WSymbol  \\
			& \cellcolor{myemph} \textbf{Evolving GNN} & \cellcolor{myemph} \WSymbol & \cellcolor{myemph} \RSymbol & \cellcolor{myemph} \RSymbol & \cellcolor{myemph} \RSymbol & \cellcolor{myemph} \WSymbol\\
			\hline
		\end{tabular}
	}
	
	\vspace{-2em}
\end{table}

\section{Preliminaries}\label{sec:prelim}
In this section, we introduce the basic concepts and formalize the graph embedding problem.
The symbols and notations frequently used throughout this paper are summarized in Table~\ref{tab:symbols}.
\begin{table}
	\centering
	\small
	\caption{ \label{tab:symbols}
		Summarization of symbols and notations.
	}
	\resizebox{\columnwidth}{!}{
		\begin{tabular}{c|c}
			\hline
			\rowcolor{mygray}
			Symbols or Notations & Description \\ \hline
			$\G$ & graph or attributed heterogeneous graph \\
			$\G^{(t)}$ & the graph at the timestamp $t$\\
			$\V$ ($\V^{(t)}$) & vertex set (at timestamp $t$)\\
			$\E$ ($\E^{(t)}$) & edge set (at timestamp $t$)\\
			$n, m$ & number of vertices and edges\\
			$\W$ ($\W^{(t)}$) & edge weight assigning function (at timestamp $t$) \\
			$\T_{V}$ & vertex type mapping function \\
			$\T_{E}$ & edge type mapping function \\
			$\A_{V}$ & vertex attributes mapping function \\
			$\A_{E}$ & edge attributes mapping function \\   
			$\bx_{v,i}$ & the $i$-th feature vector of vertex $v$ \\
			$\by_{e,i}$ & the $i$-th feature vector of edge $e$ \\ 
			$d$ & the embedding dimension \\
			$Nb(v)$ & neighbors set of vertex $v$\\
			$\bv$ & the embedding vector of vertex $v$ \\
			$\mathbf{h}_{v, c}$ &  the embedding vector of vertex $v$ w.r.t. type $c$\\
			$D^{(t)}_{i}(v)$ & number of $t$-hop in-neighbors of $v$ \\
			$D^{(t)}_{o}(v)$ & number of $t$-hop out-neighbors of $v$ \\   
			$Imp^{(t)}(v)$ & importance of vertex $v$ \\
			\hline
		\end{tabular}
	}
	\vspace{-2.5em}
\end{table}
We start with the acyclic, simple graph $\G = (\V, \E, \W)$ where $\V$ and $\E$ represent the set of vertices and edges, respectively; and $W: \E \to \mathbb{R}^{+}$ is a function assigning each edge $(u, v) \in \E$ a weight $\W(u, v)$ indicating the strength of relationships between vertex $u$ and $v$. Let $n = |\V|$ and $m = |\E|$ denote the number of vertices and edges in $\G$, respectively. Notice that, the graph $\G$ can be either directed or undirected. If $\G$ is directed, $(u, v)$ and $(v, u)$ represent two different edges and may have different weights; otherwise, $(u, v)$ and $(v, u)$ are the same edge and we have $\W(u, v) = \W(v, u)$. For each vertex $u$, we use $Nb(u)$ to denote the set of its (in and out) neighbors.

\begin{figure}[ht]
	\centering
	\includegraphics[width=0.4\textwidth]{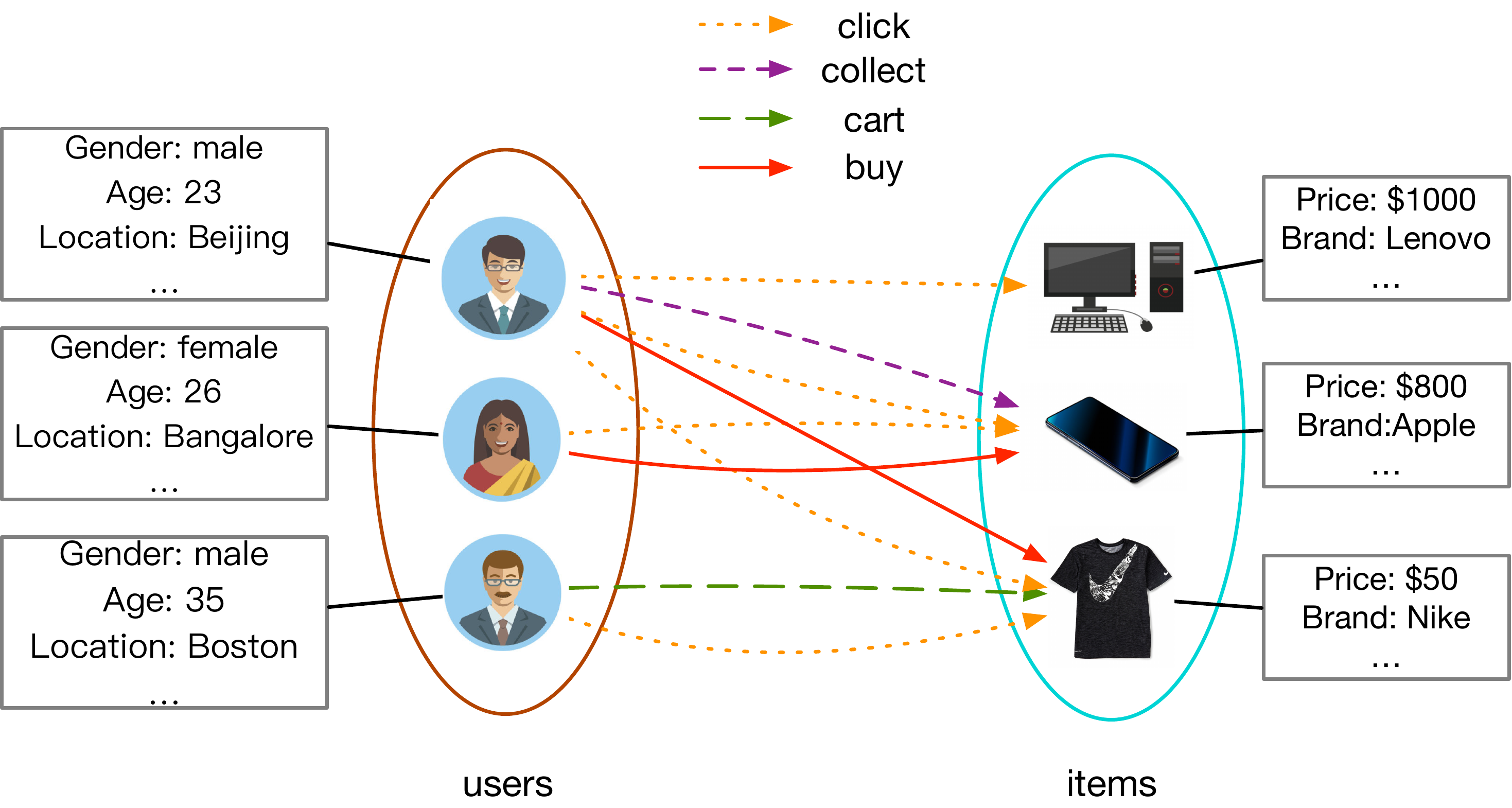}
	\vspace{-1em}
	\caption{Illustrative example of AHG with multiple types of edges, nodes and rich attributes.}
	\label{fig:AHG}
	\vspace{-1em}
\end{figure}

\smallskip
\noindent
\textbf{\underline{Attributed Heterogeneous Graph.}}
To comprehensively characterize the real-world commercial data, the practical graphs often contain rich content information, e.g., multiple types of vertices, multiple types of edges, attributes and etc. Thus, we further define the \emph{Attributed Heterogeneous Graph }(AHG). An AHG $\G$ is a tuple $(\V, \E, \W, \T_{V}, \T_{E}, \A_{V}, \A_{E})$ where $\V$, $\E$ and $\W$ have the same meaning as the simple graphs. $\T_{V}: \V \to \F_{V}$ and $\T_{E}: \E \to \F_{E}$ represent the vertex type and edge type mapping functions, where $\F_{V}$ and $\F_{E}$ are the set of vertex types and edge types, respectively. To ensure the heterogeneity, we request $|\F_{V}| \geq 2$ and/or $|\F_{E}| \geq 2$. $\A_{V}$ and $\A_{E}$ are two functions assigning each vertex $v \in \V$ and each edge $e \in \E$ some feature vectors representing its attributes. We denote the $i$-th feature vector of vertex $v$ and edge $e$ as $\bx_{v,i}$ and $\by_{e,i}$, respectively. An example of AHG is shown in 
Figure~\ref{fig:AHG}, which contains two types of vertices, namely users and items, 
and four types of edges connecting them.

\smallskip
\noindent
\textbf{\underline{Dynamic Graph.}}
Real-world graphs usually evolve with time. Given a time interval $[1, T]$, a \emph{dynamic graph} is a series of graphs $\G^{(1)}, \G^{(2)}, \break \dots, \G^{(T)}$. For each $1 \leq t \leq T$, $\G^{(t)}$ can be a simple graph or an AHG. For ease of notation, we add a superscript $(t)$ to represent the corresponding state of the objects at timestamp $t$. For example, $\V^{(t)}$ and  $\E^{(t)}$ represent the vertex set and edge set of graph $\G^{(t)}$, respectively.

\smallskip
\noindent
\textbf{\underline{Problem Definition.}}
Given an input graph $\G$, which is a simple graph or an AHG, and a predefined number $d \in \mathbb{N}$ on the dimension of embedding where $d << |\V|$, the embedding  problem is to convert the graph $\G$ into the $d$-dimensional space such that the graph property is preserved as much as possible.
GNN is a special kind of graph embedding method, which learns the embedding results by applying neural networks on graphs.
Notice that, in this paper, we concentrate on the vertex-level embedding. That is, the embedding output is a $d$-dimensional vector $\bv \in \mathbb{R}^{d}$ for each vertex $v \in \V$. 
In our future work as discussed in Section~7, we will also consider the embedding on edges, subgraphs or even the whole graph.

\section{System}\label{sec:system}

\begin{figure}[ht]
	\centering
	\includegraphics[width=0.8\columnwidth]{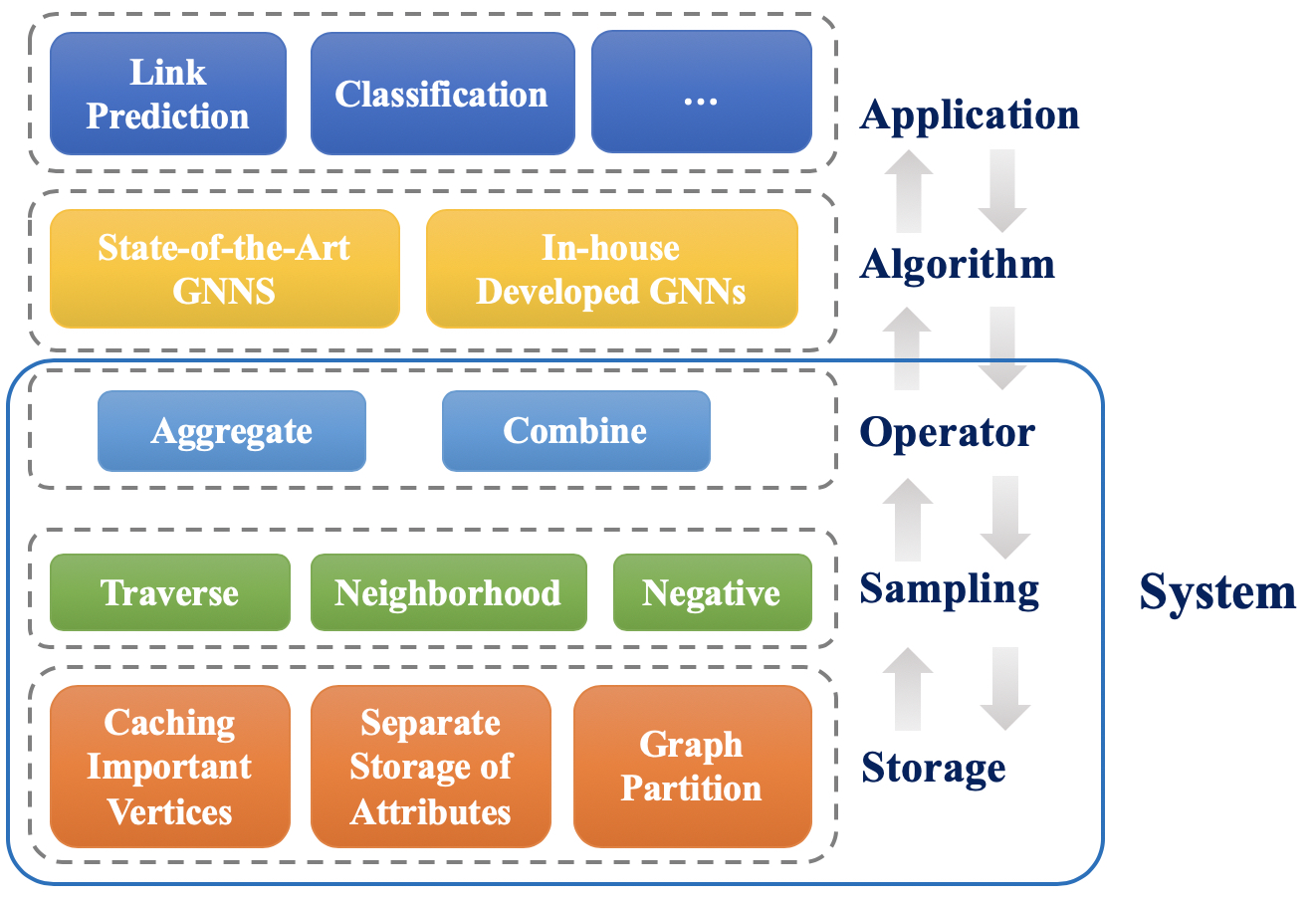}
	\vspace{-1em}
	\caption{Architecture of the AliGraph system.
	}
	\vspace{-1em}
	\label{fig:architecture}
\end{figure}

In our AliGraph platform, whose architecture is shown in Figure~\ref{fig:architecture}, we design and implement an underlying system (marked in blue square) to well support high-level GNN algorithms and applications. 
The details of this system will be described in this section. To start with, in Section~3.1, we abstract a general framework of GNNs to explain why our system is designed in this way. Sections~3.2 to 3.5 introduce the design and implementation details of each key component in the system.

\subsection{Framework of GNN Algorithms}

In this subsection, we abstract a general framework to
GNN algorithms. A series of classic GNNs such as \textsf{Structure2Vec}~\cite{Ribeiro2017struc2vec}, \textsf{GCN}~\cite{kipf2017semi}, \textsf{FastGCN}~\cite{chen2018fastgcn}, \textsf{AS-GCN}~\cite{huang2017accelerated} and \textsf{GraphSAGE}~\cite{hamilton2017inductive} can 
be characterized by instantiating the operators in the framework. The input of the GNN framework includes a graph $\G$, the embedding dimension $d \in \mathbb{N}$, a vertex feature $\bx_v$ for each vertex $v \in \V$ and the maximum hops of neighbors $k_{max} \in \mathbb{N}$.
The output of the GNN is an embedding vector $\mathbf{h}_v \in \mathbb{R}^{d}$ for each vertex $v \in \V$ and will be fed into the downstream machine learning tasks, such as classification, link prediction and etc.  

The GNN framework is described in Algorithm~\ref{algo:gnn}. At the very beginning, the vertex embedding $\mathbf{h}_v^{(0)}$ of vertex $v$ is initialized to be equal to the input attribute vector $\bx_v$. Then, at each $k$, each vertex $v$ aggregates the embeddings of its neighbors to update the embedding of itself. Specifically, we apply the \textsc{Sample} function to fetch a subset $S$ of vertices based on the neighbor set $Nb(v)$ of vertex $v$, aggregate the embeddings of all vertices $u \in S$ by the \textsc{Aggregate} function to obtain a vector $\mathbf{h}_v'$, and combine $\mathbf{h}_v'$ with $\mathbf{h}_v^{(k-1)}$ to generate the embedding vector  $\mathbf{h}_v^{(k)}$ by the \textsc{Combine} function. After processing all vertices, the embedding vectors are normalized. Finally, after $k_{max}$ hops, $\mathbf{h}_v^{(k_{max})}$ is returned as the embedding result $\mathbf{h}_v$ of vertex $v$. 

\begin{algorithm}[t]
	\scriptsize
	\caption{\textsf{GNN Framework} \label{algo:gnn}}
	\KwIn{network $\mathcal{G}$, embedding dimension $d \in \mathbb{N}$, a vertex feature $\bx_v$ for each vertex $v \in \V$ and the maximum hops of neighbors $k_{max} \in \mathbb{N}$.}
	\KwOut{embedding result $\mathbf{h}_v$ of each vertex $v \in \V$}
	$\mathbf{h}_v^{(0)} \gets \bx_v$\\
	\For{$k \gets 1$ to $k_{max}$}{
		\For{each vertex $v \in \V$}{
			$S_v \gets \textsc{Sample}(Nb(v))$\\ 
			$\mathbf{h}'_v \gets \textsc{Aggregate}(\mathbf{h}_{u}^{(k-1)}, \forall u \in S)$\\
			$\mathbf{h}_{v}^{(k)} \gets \textsc{Combine}(\mathbf{h}_{v}^{(k-1)}, \mathbf{h}'_v)$\\
		}
		normalize all embedding vectors $\mathbf{h}_v^{(k)}$ for all $v \in \V$\\
	}
	$\mathbf{h}_v \gets \mathbf{h}_v^{(k_{max})}$ for all $v \in \V$
	return $\mathbf{h}_v$ as the embedding result for all $v \in \V$
\end{algorithm}

\smallskip
\noindent{\underline{\textbf{System Architecture.}}}
Based on the GNN framework described above, we naturally construct the system architecture of the AliGraph platform, as shown in Figure~\ref{fig:architecture}. Notice that, the platform consists of five layers on the whole, where the three underlying layers form the system to support the \textit{algorithm} layer and the \textit{application} layer.
Inside the system, the \textit{storage} layer organizes and stores different kinds of raw data to fulfill the fast data access requirements of high-level operations and algorithms.
Upon this, by Algorithm~\ref{algo:gnn}, we find that three main operators, namely \textsc{Sample}, \textsc{Aggregate} and \textsc{Combine}, play important roles in various GNN algorithms.
Among them, the \textsc{Sample} operator lays foundation for
\textsc{Aggregate} and \textsc{Combine} since it directly controls the scope of information to be processed by them.
Therefore, we design the \textit{sampling} layer to access the storage for fast and accurate generation of training samples. Above it, the \textit{operator} layer specifically optimizes the \textsc{Aggregate} and \textsc{Combine} functions. 
On top of the system, the GNN algorithms can be constructed in the \textit{algorithm} layer to serve real-world tasks in the \textit{application} layer.

\subsection{Storage}
In this subsection, we discuss how to store and organize the raw data. Notice that, the space cost to store the real-world graphs is very large. Common e-commerce graphs can contain tens of billions of nodes and hundreds of billions of edges  with storage cost over 10TB easily. The large graph size brings great challenges for efficient graph access, especially in a distributed environment of clusters. To well support the high-level operators and algorithms, we apply the following three strategies in the \textit{storage} layer of \textit{AliGraph}.

\smallskip
\noindent{\underline{\textbf{Graph Partition.}}}
Our AliGraph platform is build on a distributed environment, thus the whole graph is divided and separately stored in different workers. The goal of graph partition is to minimize the number of crossing edges whose endpoints are in different workers. To this end, literature work has proposed a series of algorithms. In our system, as recommended in~\cite{fan2017grape}, we implement four built-in graph partition
algorithms: 1) METIS~\cite{KarypisMETIS};
2) Vertex cut and edge cut partitions~\cite{Gonzalez2012PowerGraph};
3) 2-D partition~\cite{Boman2013Scalable}; 
and 4) Streaming-style partition strategy~\cite{Stanton2013Streaming}.
These four algorithms are suitable to different circumstances. 
In short, the METIS method is specialized in processing sparse graphs;
the vertex and edge cut method performs much better on dense graphs;
2-D partition is often used when the number of workers is fixed;
and streaming-style partition method are often applied on graphs with frequently edge updates.
Users can choose the best partition strategy based on their own needs, moreover, they can also implement other graph partition algorithms as plugins in the system.

In Algorithm~\ref{algo:partition}, lines~1--4 present the interface of graph partition. For each edge $e \in \E$, the general function \textsc{Assign} in line~4 computes which worker $e$ will be in based on its endpoints.

\smallskip
\noindent{\underline{\textbf{Separate Storage of Attributes.}}}
Notice that, for AHGs, we need to store both the structural and attributes of the partitioned graphs in each worker.
The structural information of graph can be simply stored by an adjacency table. That is, for each vertex $v$, we store its neighbor set $Nb(v)$. Whereas, for the attributes on both vertices and edges, it is inadvisable to store them together in the adjacency table. The reasons are two-fold:
1) Attributes often cost more spaces. For example, the space cost to store a vertex id is at most $8$ bytes while the attributes on a vertex may range from $0.1$KB to $1$KB.
2) Attributes among different vertices or edges have largely overlaps. For example, many vertices may have the same tag ``man'' indicating its gender. 
Therefore, it is more reasonable to separately store attributes.

\begin{figure}[t]
	\centering
	\includegraphics[width=0.4\textwidth]{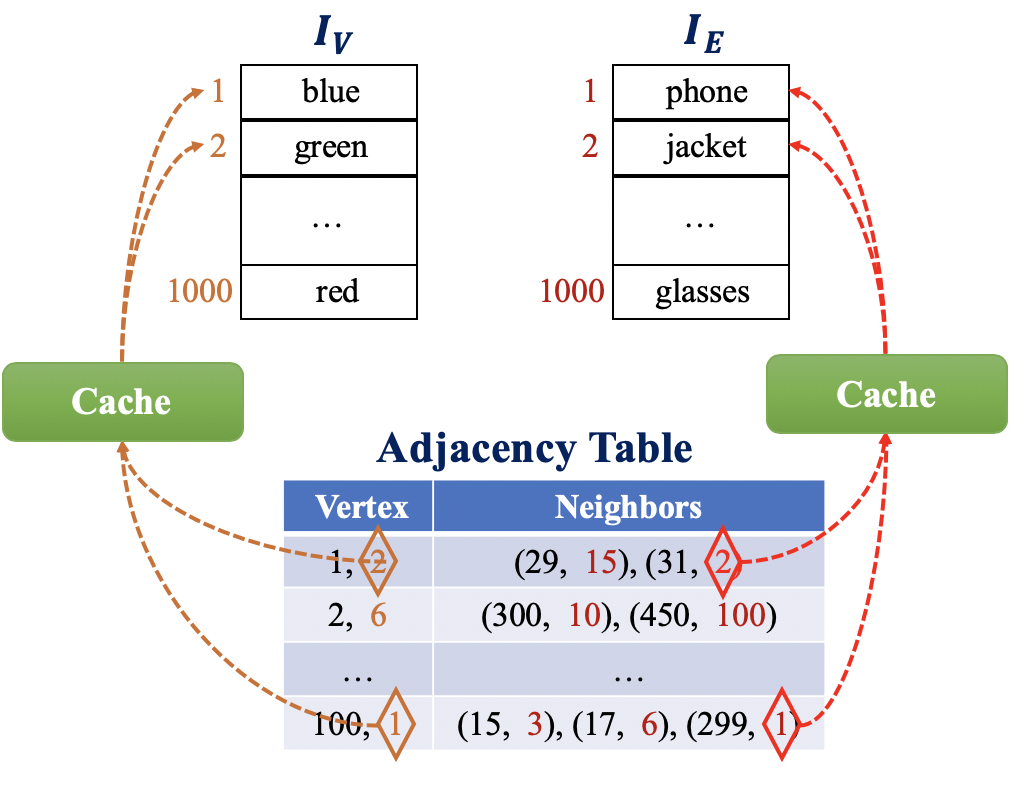}
	\vspace{-2em}
	\caption{Index structure of graph storage.}
	\label{fig:index}
	\vspace{-2em}
\end{figure}

In our system, we do so by building two indices 
$I_{\V}$ and $I_{\E}$ to store the attributes on vertices and edges, respectively. Each entry in 
$I_{\V}$ ($I_{\E}$ resp.) is a unique attribute  associated on vertex (edge resp.). 
As illustrated in Figure~\ref{fig:index}, in the adjacency table, for each vertex $u$, we store the index of attribute $\A_{\V}(u)$ in $I_{\V}$, and for each edge $(u, v)$, we also store the index of attribute $\A_{\E}(u, v)$  in $I_{\E}$. Let $N_D$ and $N_L$ be the average number of neighbors and average length of attributes. Let $N_{A}$ be the number of distinct attributes on vertices and edges. Obviously, our separate storage strategy decreases the space cost from $O(n N_{D} N_{L})$ to $O(nN_{D} + N_{A}N_{L})$.

Undoubtedly, separate storage of the attributes will increase the access time for retrieving the attributes. On average, each vertex will need to access the index $I_{\E}$ at most $N_D$ times to collects the attributes of all of its neighbors. To mitigate this, we add two cache components to reside the frequently accessed items in $I_{\V}$ and $I_{\E}$, respectively. We adopt the least recently used (LRU) replacing strategy~\cite{Chrobak1998LRU} in each cache.

\smallskip
\noindent{\underline{\textbf{Caching Neighbors of Important Vertices.}}}
In each worker, we further propose a method to locally cache the neighbors of some important vertices to reduce the communication cost. The intuitive idea is that if a vertex $v$ is frequently accessed by other vertices, we can store $v$'s out-neighbors in each partition it occurs. By doing this, the visiting cost of other vertices to their neighbors via $v$ can be greatly reduced. However, if the number of neighbors of $v$ is large, storing multiple copies of $v$'s neighbors will also incur huge storage cost. To make a better trade-off, we define a metric to evaluate the \emph{importance} of each vertex, which decides whether a vertex is worth to cache or not. 

Let $D^{(k)}_{i}(v)$ and $D^{(k)}_{o}(v)$ denote the number of $k$-hop in and out-neighbors of the vertex $v$, respectively. Certainly, $D^{(k)}_{i}(v)$ and $D^{(k)}_{o}(v)$ can measure the benefit and cost of caching the out-neighbors of $v$, respectively. Thus, the $k$-th \textit{importance} of $v$, denoted as $Imp^{(k)}(v)$, is defined as
\begin{equation}
\label{eq:Impkv}
Imp^{(k)}(v) = \frac{D^{(k)}_{i}(v)}{D^{(k)}_{o}(v)}.
\end{equation}
We only cache the out-neighbors of a vertex $v$ if its importance value $Imp^{(k)}(v)$ is sufficiently large. In Algorithm~\ref{algo:partition}, lines~5--9 present the process of caching neighbors of important vertices. Let $h$ denote the maximum depth of neighbors we consider. For each vertex $v$, we cache the $1$ to $k$-hop out-neighbors of $v$ if $Imp^{(k)}(v) \geq \tau_{k}$, where $\tau_{k}$ is a user-specified threshold. Notice that, setting $h$ to a small number, usually $2$, is enough to support a series of practical GNN algorithms. Practically, we find that  $\tau_{k}$ is not a sensitive parameter. By experimental evaluation, setting $\tau_{k}$ to a small value around $0.2$ can make the best trade-off between cache cost and benefit. 

Interestingly, we find that the vertices to be cached is only a very small part of the whole graph. As analyzed in~\cite{Tanimoto2009Power}, the direct in and out-degree of vertices in real-world graphs, i.e., $D^{(1)}_{i}(v)$ and $D^{(1)}_{o}(v)$, often obey the power-law distribution. That is, only a very few vertices in the graph have large in and out-degree. Based on this, we derive the following two theorems. The proof can be found in the appendix.

\begin{theorem}
	\label{Thm:tHopPowerlaw}
	If the in and out-degree distribution of the graph obey the power-law distribution, for any $k \geq 1$, the number of $k$-hop in and out-neighbors of the vertices in the graph also obey the power-law distribution.
\end{theorem}

\begin{theorem}
	\label{Thm:ImpPowerlaw}
	If the in and out-degree distribution of the graph obey the power-law distribution, the importance value of the vertices in the graph also obey the power-law distribution.
\end{theorem}

Theorem~\ref{Thm:ImpPowerlaw} indicates that only a very few vertices in the graph have large importance values. That means, we only need to cache a small number of important vertices to achieve a significant cost decrease  of graph traversals.

\begin{algorithm}
	\scriptsize
	\caption{\textsf{Partition and Caching} \label{algo:partition}}
	\KwIn{graph $\G$, partition number $p$,
		cache depth $h$, threshold $\tau_1, \tau_2, \dots, \tau_h$}
	\KwOut{$p$ subgraphs}
	Initialize $p$ graph servers
	
	\For{each edge $e = (u, v) \in \E$} {
		$j = \textsc{Assign}(u)$ \\
		Send edge $e$ to the $j$-th partition \\
	}
	\For{each vertex $v \in V$}
	{
		\For{$k \gets 1 \text{ to } h$}
		{
			Compute $D_{i}^{(k)}(v)$ and $D_{o}^{(k)}(v)$ \\
			\If{$\frac{D_{i}^{(k)}(v)}{D_{o}^{(k)}(v)}\geq \tau_k$}
			{
				Cache the $1$ to $k$-hop out-neighbors of $v$ on each partition where $v$ exists
			}
		}
	}
\end{algorithm}

\subsection{Sampling}

Recall that, GNN algorithms rely on aggregating neighborhood information to generate embeddings of each vertex. However, the degree distribution of real-world graphs is often skewed~\cite{Tanimoto2009Power}, which makes the convolution operation hard to operate.
To tackle this, existing GNNs usually adopt various sampling strategies to sample a subset of neighbors with aligned sizes. Due to its importance, in our system, we abstract a sampling layer specified to optimize the sampling strategies.

\smallskip
\noindent{\underline{\textbf{Abstraction.}}}
Formally, the sampling function takes input a vertex subset $V_T$ and extracts a small subset $V_S \subseteq V_T$ such that $|V_S| << |V_{T}|$. By taking a thorough overview of current GNN models, we abstract three kinds of different samplers, namely \textsc{Traverse}, \textsc{Neighborhood} and \textsc{Negative}. 

\begin{itemize}
	\item \textsc{Traverse}:
	is used to sampling a batch of vertices or edges from the whole partitioned subgraphs.
	
	\item \textsc{Neighborhood}:
	will generate the context for a vertex. The context of this vertex may be one or multi hop neighbors, which are used to encode this vertex.
	
	\item \textsc{Negative}: 
	is used to generate negative samples to accelerate the convergence of the training process.
\end{itemize}

\smallskip
\noindent{\underline{\textbf{Implementation.}}}
In the literature, the sampling method plays an important role to enhance the efficiency and accuracy of the GNN algorithms~\cite{GraphSage:HamiltonYL17, Hamilton2017Representation, Cai2017A, Goyal2018Graph}. In our system, we treat all samplers as plugins. Each of them can be implemented independently. The three types of samplers can be implemented as follows.

For \textsc{Traverse} samplers, they  get data from the local subgraphs. For \textsc{Neighborhood} samplers, they can get one-hop neighbors from local storage as well as multi-hop neighbors from local cache. If the neighbors of a vertex are not cached, a call to remote graph server is needed. When getting the context of a batch of vertices, we first partition the vertices into sub-batches, and the context of each sub-batch will be stitched together after being returned from the corresponding graph server. \textsc{Negative} samplers usually generate samples from local graph server. For some special cases, negative sampling from other graph server may be needed. Negative sampling is flexible in algorithm, and we do not need to call all graph servers in a batch. In summary, a typical sampling stage can be achieved as illustrated in Figure~\ref{fig:sampling}.

\begin{figure}[ht]
	\scriptsize
	\centering
	\fbox{\parbox{0.8\columnwidth}{
			\texttt{Define a TRAVERSE sampler as s1}\\
			\texttt{Define a NEIGHBORHOOD sampler as s2}\\
			\texttt{Define a NEGATIVE sampler as s3}\\
			\textbf{...} \\
			\textbf{\color{brown}{def}} \textbf{ sampling(s1, s2, s3, batch\_size):} \\
			\text{\quad vertex = s1.sample(edge\_type, batch\_size)} \\
			\text{\quad \color{gray}{\# hop\_nums contains neighbor count at each hop}} \\
			\text{\quad context = s2.sample(edge\_type, vertex, hop\_nums)} \\
			\text{\quad neg = s3.sample(edge\_type, vertex, neg\_num)} \\
			\textbf{\color{blue}{\quad return}} \text{ vertex, context, neg} \\ 
	}}
	\caption{Sampling stage using three kinds of samplers}
	\vspace{-1em}
	\label{fig:sampling}
\end{figure}

We can accelerate training by adopting several efficient sampling strategies with dynamic weights.  We implement the update operation in a sampler's backward computation, just like gradient back propagation~\cite{Hush1988Improving}  of an operator. So when updating needed, what we should do is to register a gradient function for the sampler. The updating mode, synchronous or asynchronous, is due to the training algorithm.

\begin{figure}[ht]
	\centering
	\includegraphics[width=0.4\textwidth]{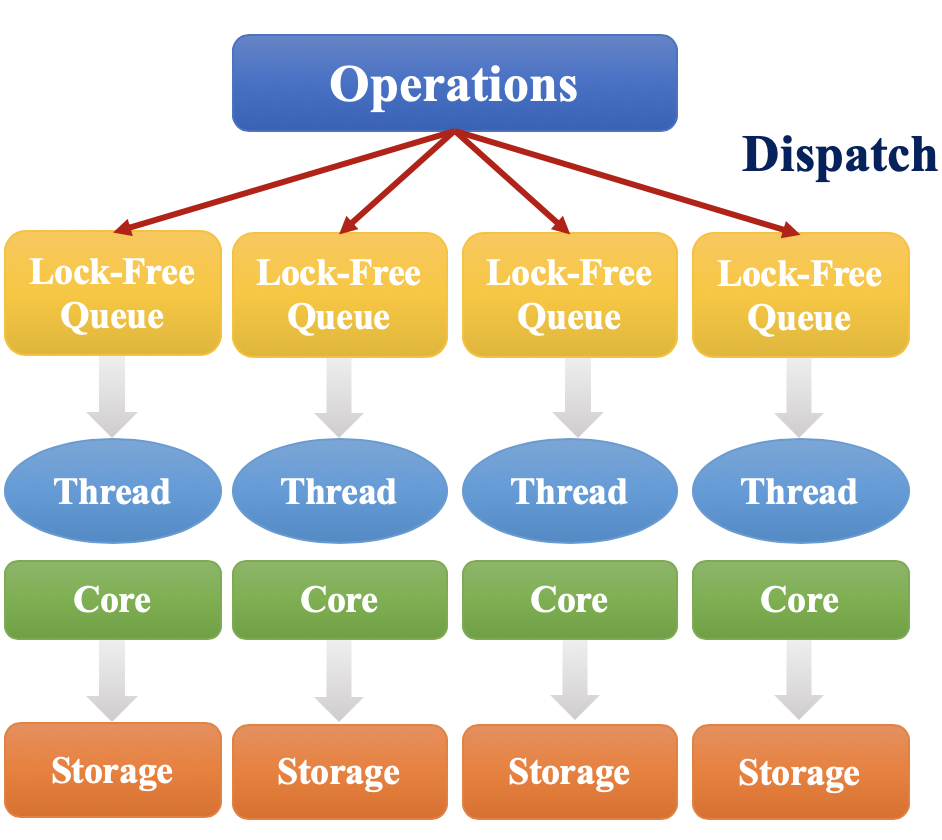}
	\vspace{-1em}
	\caption{Lock-free graph operations}
	\label{fig:graph_op}
	\vspace{-2em}
\end{figure}

Till now, both reading and updating will be operated on the graph storage in memory, which may lead to weak performance. According to the neighborhood requirement, the graph is partitioned by source vertices. Based on this, we split the vertices on a graph server into groups. Each group will be related with a request-flow bucket, in which the operations, including reading and updating, are all about the vertices in this group. The bucket is a lock-free queue. As shown in Figure~\ref{fig:graph_op}, we bind each bucket to a CPU core and then each operation in the bucket will be processed sequentially without locking, which will further enhance the efficiency of the system.

\subsection{Operator}

\noindent{\underline{\textbf{Abstraction.}}}
After sampling, the output data is aligned, and we can process it easily. Upon samplers, we need some GNN-like operators to consume them. In our system, we abstract two kinds of operators, namely \textsc{Aggregate} and \textsc{Combine}~\cite{kipf2017semi,chen2018fastgcn,huang2017accelerated,GraphSage:HamiltonYL17}. Their roles are as follows.

\begin{itemize}
	\item \textsc{Aggregate}: 
	collects the information of each vertex's neighbors to produce a unified result. For example, the \textsc{Aggregate} function in Algorithm~\ref{algo:gnn} maps a series of vectors $\mathbf{h}_u \in \mathbb{R}^{d}$ to a single vector $\mathbf{h}_v' \in \mathbb{R}^{d}$, where $u$ belongs to sampled neighborhood nodes of $v$. $\mathbf{h}_v'$ is an intermediate result to further generate $\mathbf{h}_v^{(k)}$.
	The \textsc{Aggregate} function acts as the \emph{convolution} operation since it collects the information from its surrounding neighborhoods. In different GNN methods, a variety of aggregating methods are applied, such as element-wise mean, max-pooling neural network and long short-term memory (LSTMs)~\cite{GraphSage:HamiltonYL17,Hamilton2017Representation}. 
	
	\item \textsc{Combine}: takes care of how to use neighbors of a vertex to describe the vertex. In Algorithm~\ref{algo:gnn}, the \textsc{Combine} function maps the two vectors $\mathbf{h}_{v}^{(k-1)}$ and $\mathbf{h}'_v \in \mathbb{R}^{d}$ into a single vector $\mathbf{h}^{(k)}_v \in \mathbb{R}^{d}$. The \textsc{Combine} function can integrate the information of the previous hop and the neighborhoods into an unified space. Usually, in existing GNN methods, $\mathbf{h}_{v}^{(k-1)}$ and $\mathbf{h}'_v$ are summed together to fed into a deep neural network.
\end{itemize}

\smallskip
\noindent{\underline{\textbf{Implementation.}}}
Notice that, both samplers and GNN-like operators not only do computations forward, but also take charge of parameters updating backward if needed. So that we can make the whole model as a network for an end-to-end training. Considering the characteristics of graph data, a lot of optimization can be taken into account to achieve better performances. Similar to \textsc{Sample}, \textsc{Aggregate} and \textsc{Combine} are plugins of \textit{AliGraph}, which can be implemented independently. A typical operator is made up of forward and backward computations to be easy to be involved in a deep network. Based on operators, users can set up a GNN algorithm quickly.

To further accelerate the computation of the two operators, we applying  strategies by materialization of intermediate vectors $\mathbf{h}_v^{(k)}$. Notice that, 
as shown in~\cite{chen2018fastgcn}, in each mini-batch during the training process, we can share the set of sampled neighbors for all vertices in the mini-batch. As well, we can also share the vectors $\mathbf{h}_v^{(k)}$ for all $1 \leq k \leq k_{max}$ among vertices in the same mini-batch. To this end, we store $k_{max}$ vectors $\mathbf{\hat{h}}^{(1)}_v, \mathbf{\hat{h}}^{(2)}_v, \dots, \mathbf{\hat{h}}^{(k_{max})}_v$ to be the newest  vectors of all vertices in the mini-batch. 
Then, in the \textsc{Aggregate} function, we apply vectors in $\mathbf{\hat{h}}^{(1)}_v, \mathbf{\hat{h}}^{(2)}_v, \dots, \mathbf{\hat{h}}^{(k_{max})}_v$ to obtain $\mathbf{\hat{h}}_v'$. After that, we apply $\mathbf{\hat{h}}_v'$ and $\mathbf{h}_v^{(k-1)}$ to compute $\mathbf{\hat{h}}_v^{(k)}$ by the \textsc{Combine} function. Finally, the stored vector $\mathbf{\hat{h}}^{(k)}$ is updated by $\mathbf{\hat{h}}_v^{(k)}$. By this strategy, the computation cost on the operators can be greatly reduced.

\section{Methodology}\label{sec:methodology}
On top of the system, we discuss the design of algorithms in this section. We show that existing GNNs can be easily built on \textit{AliGraph}. Besides, we also propose a bunch of new GNN algorithms to tackle the four newly arisen challenges of embedding real-world graph data as summarized in Section 1.  All of them are plugins in the \emph{algorithm} layer of the AliGraph platform.

\subsection{State-of-the-Art GNNs}
As our AliGraph platform is abstracted from upon the general GNN algorithms, existing GNNs can be easily implemented on this platform. Specifically, 
the GNNs listed in Table~\ref{tab:methods}
can all be built in AliGraph by following the framework in Algorithm~\ref{algo:gnn}. Here we take the \textsf{GraphSAGE}  as an example. Other GNNs can be implemented in a similar way. We omit them due to space limitations. Notice that, for \textsf{GraphSAGE}, it applies a simple node-wise sampling to extract a small subset from the neighbor set of each vertex. Obviously, its sampling strategy can be easily implemented by using our \textsc{Sampling} operator. Then, we need to instantiate the \textsc{Aggregate} and \textsc{Combine} functions in Algorithm~\ref{algo:gnn}. 
The \textsf{GraphSAGE} can apply the weighted element-wise mean in the \textsc{Aggregate} function in line~4. 
Besides, other more complex functions such as the max-pooling neural network and LSTM can also be used. 
In other GNN methods such as \textsf{GCN}, \textsf{FastGCN} and \textsf{AS-GCN}, we can replace different strategies on   \textsc{Sampling}, \textsc{Aggregate} and \textsc{Combine}.

\subsection{In-House Developed GNNs}
Our in-house developed GNNs focus on various aspects, e.g., sampling (\textsf{AHEP}), multiplex (\textsf{GATNE}),  multimode (\textsf{Mixture GNN}), hierarchy (\textsf{Hierarchical GNN}), dynamic (\textsf{Evolving GNN}) and multi-sourced information (\textsf{Bayesian GNN}). 

\medskip
\noindent{\underline{\textbf{\textsf{AHEP} Algorithm.}}}
This algorithm is designed to mitigate the heavy computation and storage costs of the traditional embedding propagation (\textsf{EP}) algorithm ~\cite{duran2017learning} on heterogeneous networks, \textsf{HEP}~\cite{zheng2018heterogeneous}. \textsf{HEP} follows the general framework of GNN with minor modifications adapted to AHG. 
Specifically, in \textsf{HEP}, the embeddings of all vertices are generated in an iterative manner. In the $k$-th hop, for each vertex $v$ and each node-type $c$, all neighbors $u$ of $v$ in type $c$ propagate its embedding $\bm{h}_{u, c}$ to $v$ to reconstruct an embedding $\bm{h}'_{v, c}$. The embedding of $v$ is then updated by concatenating $\bm{h}'_{v, c}$ across all node types.
Whereas, in \textsf{AHEP} (\textsf{HEP} with adaptive sampling), we sample important neighbors instead of considering the whole set of neighbors. During this process, we design a metric to evaluate the importance of each vertex by incorporating its structural information and features. After that, all neighbors in different types are separately sampled according to their corresponding probability distributions. We carefully design the probability distributions to minimize the sampling variance. In a specific task, to optimize the \textsf{AHEP} algorithm, the loss function can be generally described 
as 
\begin{equation}
L = L_{SL} + \alpha L_{EP} + \beta \Omega(\bm{\Theta}),
\end{equation}
where $L_{SL}$ is the loss from supervised learning in the batch, $L_{EP}$ is the embedding propagation loss with sampling in the batch, $\Omega(\bm{\Theta})$ is the regularizer of all trainable parameters, and $\alpha, \beta$ are two hyperparameters. As verified by experimental results in Section~5, \textsf{AHEP} runs much faster than \textsf{HEP} while achieving comparable accuracies.  

\medskip
\noindent{\underline{\textbf{\textsf{GATNE} Algorithm.}}}
This algorithm is designed to cope with graphs with heterogeneous and attribute information on both vertices and edges. To address the above challenges, we propose a novel approach to capture both rich attributed information and to utilize multiplex topological structures from different node types, namely  \textbf{G}eneral \textbf{A}ttributed Multiplex  He\textbf{T}erogeneous \textbf{N}etwork \textbf{E}mbedding, or abbreviated as \textsf{GATNE}. The overall embedding result of each vertex consists of three parts: the general embedding, the specific embedding and the attribute embedding, which correspondingly characterize the structural information, the heterogeneous information and the attribute information, respectively. For each vertex $v$ and any node type $c$, the general embedding $\mathbf{b}_v$ and the attribute embedding $\mathbf{f}_v$ keep the same. Let $t$ be an adjustable hyper-parameter and $\mathbf{g}_{v, t'}$ where $1\leq t'\leq t$ be meta-specific embeddings. The specific embedding $\mathbf{g}_v$ is obtained by concatenating all $\mathbf{g}_{v, t'}$. Then, for each type $c$, the overall embedding of $\mathbf{h}_{v, c}$ w.r.t.~$c$ can be written as

\begin{equation}
\label{eqn:vi}
\mathbf{h}_{v, c} = \mathbf{b}_v + \alpha_c \mathbf{M}_c^{T} \mathbf{g}_v \mathbf{a}_c + \beta_c \mathbf{D}^T \mathbf{x}_v,
\end{equation}
where $\alpha_c$ and $\beta_c$ are two adjustable coefficients reflecting the importance of the specific embedding and the attribute embedding; the matrix $\mathbf{a}_c \in \mathbb{R}^m$ of coefficients are computed by using the self-attention mechanism in~\cite{lin2017structured}; and $\mathbf{M}_c$ and $\mathbf{D}$ are two trainable transformation matrices.
The final embedding result $\mathbf{h}_v$ can then be obtained by concatenating all $\mathbf{h}_{v, c}$.

The embeddings can be learned by applying the random walk based methods similar to~\cite{perozzi2014deepwalk,grover2016node2vec}. Specifically, given a vertex $v$ in type $c$ in a random walk and the window size $p$, let $v^{-p}, v^{-p+1}, \dots,\break v, v^{1}, \dots, v^{p}$ denote its context. We need to minimize the negative log-likelihood as 

\begin{equation}
\label{eqn:obj_fun}
-\log P_{\theta_c} \left(v^{-p},\ldots,v^{p}|v \right)= \!\!\! \sum_{1 \leq |p'|\leq p} - \log P_{\theta^c}(v_{p'}|v), 
\end{equation}
where $\theta_c$ denotes all the parameters w.r.t.~type $c$ and $P_{\theta_c}(v_{p'}|v)$ is defined by the softmax function. The objective function $-\log P_{\theta_c}(u|v)$ for each pair of vertices $u$ and $v$ can be easily approximated by the negative sampling method.

\medskip
\noindent{\underline{\textbf{\textsf{Mixture GNN}.}}}
This model is a mixture GNN model to tackle with the heterogeneous graphs with multi-modes. In this model, we extend the skip-gram model on homogeneous graphs~\cite{perozzi2014deepwalk} to fit the polysemous situation on heterogeneous graphs. In the traditional skip-gram model, we try to find the
embedding of graphs with parameters $\theta$ through maximizing the likelihood as
\begin{equation}\label{eq:overall}
L_\theta  = \log \Pr\nolimits_{\theta}(Nb(v)|v) = \sum_{u \in Nb(v)} \log \Pr\nolimits_{\theta}(u|v),
\end{equation}
where $Nb(v)$ denotes the neighbors of $v$ and $\Pr\nolimits_{\theta}(u|v)$ is a softmax function.
In our setting on heterogeneous graphs, each node owns multiple senses. To differentiate them, let $P$ be the known distribution of node senses. We can rewrite the objective function as 
\begin{equation}\label{eq:overallpoly}
L_{poly, \theta}  = \log \Pr\nolimits_{P, \theta}(Nb(v)|v) = \sum_{u \in Nb(v)} \log \Pr\nolimits_{P, \theta}(u|v).
\end{equation}
At this time, it is hard to incorporate the negative sampling metod to directly optimize Equation~\eqref{eq:overallpoly}. Alternatively, we derive a novel lower bound $L_{low}$ of Equation~\eqref{eq:overallpoly} and try to maximize $L_{low}$. We find that the terms in the lower bound $L_{low}$ can be approximated by the negative sampling. As a result, the training process can be easily   implemented by slightly modifying the sampling process in existing work such as Deepwalk~\cite{perozzi2014deepwalk} and node2vec~\cite{grover2016node2vec}. 

\medskip
\noindent{\underline{\textbf{\textsf{Hierarchical GNN}.}}}
Current GNN methods are inherently flat and do not learn hierarchical representations of graphs: a limitation that is especially problematic to explicitly investigate such similarities of various types of user behaviors. This model combines the hierarchical structure to strengthen the expression power of GNN. Let $\BH^{(k)} \in \mathbb{R}^{|V| \times d}$ denote the matrix of node embeddings computed after $k$ steps of the GNN and $\A$ be the adjacency matrix of the graph $\G$. In Algorithm~\ref{algo:gnn}, traditional GNN iteratively learns $\BH^{(k)}$ by combining 
$\BH^{(k-1)}$, $\A$ and some trainable parameters 
$\theta^{(k)}$. Initially, we have $\BH^{(0)} = \BX$, where $\BX$ represent matrix of the node features.

In our hierarchical GNN, we learn the embedding result in a layer-to-layer fashion. Specifically, let $\A^{(l)}$ and $\BX^{(l)}$ denote the adjacency matrix and the node feature matrix in the $l$-th layer, respectively. The vertex embedding result matrix $\BZ^{(l)}$ in the $l$-th layer is learned by feeding $\A^{(l)}$ and $\BX^{(l)}$ into the single-layer GNN method. After that, we cluster some vertices in the graph and update the adjacency matrix $\A^{l}$ to $\A^{l+1}$. Let $\BS^{(l)}$ denote the learned assignment matrix in the $l$-th layer. Each row and column in $\BS^{(l)}$ corresponds to a cluster in the $l$-th and $(l+1)$-th layer, respectively. $\BS^{(l)}$ can be obtained by a softmax function applied on another pooling GNN upon $\A^{(l)}$ and $\BX^{(l)}$.
Taking $\BZ^{(l)}$ and $\BS^{(l)}$ in hand, we can obtain the new coarsened adjacency matrix $\A^{(l+1)} = {\BS^{(l)}}^T \A^{(l)} \BS^{(l)}$ and the new feature matrix $\BX^{(l+1)}={\BS^{(l)}}^T \BZ^{(l)}$ for the next $(l+1)$-th layer. As verified in Section~5, the multi-layer hierarchical GNN is more effective than the single-layer traditional GNNs.

\medskip
\noindent{\underline{\textbf{\textsf{Evolving GNN}.}}}
This model is proposed to embedding vertices in the dynamic network setting. Our goal is to learn the representations of vertices in a sequential of graphs $\G^{(1)}, \G^{(2)},..., \G^{(T)}$. To capture the evolving nature of dynamic graphs, we
divide the evolving links into two types:
1) the normal evolution representing the majority of reasonable changes of edges; and
2) burst links representing rare and abnormal evolving edges. Based on them, the embedding of all vertices in the dynamic graphs are learned in an interleave manner. Specifically, at timestamp $t$, the normal and burst links found on graph $\G^{(t)}$ are integrated with the GraphSAGE model~\cite{GraphSage:HamiltonYL17} to generate embedding results $\mathbf{h}_v$ of each vertex $v$ in $\G^{(t)}$. Then, we apply a method to predict the normal and burst information on the graph $\G^{(t+1)}$ by using Variational Autoencoder and RNN model~\cite{Kingma2013vae}.  This process is executed in iterations to output the embedding results of each vertex $v$ at each timestamp $t$.

\medskip
\noindent{\underline{\textbf{\textsf{Bayesian GNN}.}}}
This model integrates two sources of information, knowledge graph embedding (e.g., symbolic) or behavior graph embedding (e.g., relations), through the Bayesian framework.  To be more specific, it mimics the human understanding process in the cognitive science, in which each cognition is driven by adjusting the prior knowledge under a specific task. Specifically, given a knowledge graph $\G$ and an entity (vertex) $v$ in $\G$, its basic embedding $\bh_v$ is learned by purely considering $\G$ itself, which characterizes the prior knowledge in $\G$.
Then, a task-specific embedding $\bz_v$ is generated according to $\bh_v$ and a correction term $\bdelta_v$ respect to the task. That is,
\begin{equation}
\bz_v \approx \bff(\bh_v + \bdelta_v),
\label{eq:connection_z_h_1st}
\end{equation}
where $\bff$ is a non-linear function that projects $\bh_v + \bdelta_v$ as $\bz_v$. 

Notice that, learning exact $\bdelta_v$ and $\bff$ seems infeasible since each entity $v$ has a different $\bdelta_v$ and the $\bff$ function is very complex. To address this problem, we apply a generation model from $\bh_v$ to $\bz_v$ by considering the second-order information. Specifically, for each entity $v$, we sample its correction variable $\bdelta_v$ from a Guassian distribution $N(\mathbf{0}, s_v^2)$, where $s_v$ is determined by the coefficients of $\bh_v$. Then, for each pair of entities $v_1$ and $v_2$, we sample $\bold z_{v_1} - \bold z_{v_2}$ according to another Guassian distribution
$$N\left ( \bff_{\phi}(\bh_{v_1} +  \bdelta_{v_1}) - \bff_{\phi}(\bh_{v_2} + \bdelta_{v_2}), \text{diag}(\bsigma^2_{v_1} + \bsigma^2_{v_2}) \right),$$ where $\phi$ representing the trainable parameters for the function $\bff$. Let the posterior mean of $\bdelta_v$ be $\hat{\bmu}_{v}$ and $\hat{\phi}$ be the resulting parameters, we finally apply $\bh_v +\hat{\bmu}_{v}$ as the corrected embedding for the knowledge graph, and $\bff_{\hat{\phi}}(\bh_v +\hat{\bmu}_{v})$ as the corrected task-specific embedding.

\section{Experiments}\label{sec:experiments}

We conducted extensive experiments to evaluate our AliGraph platform, including both system and algorithms.  

\subsection{System Evaluation}
In this subsection, we evaluate the performance of the underlying system in the AliGraph platform from the perspectives of \textit{storage} (graph building and caching neighbors), \textit{sampling} and \textit{operator}. All experiments are carried on two datasets \textsl{Taobao-small} and \textsl{Taobao-large} described in Table~\ref{tab:system_datasets}, where the
storage size of the latter is six times larger.  Both of them represent the subgraphs of users and items 
extracted from the Taobao e-commerce Platform. 

\begin{table}
	\centering
	\caption{\label{tab:system_datasets} Datasets used in system experiments. \textsl{Taobao-large} is six times large than \textsl{Taobao-small}.
	}
	\resizebox{\columnwidth}{!}{
		\begin{tabular}{c|cccccc}
			\hline 
			\rowcolor{mygray}
			Dataset & \tabincell{c}{\# user \\ vertices} & 
			\tabincell{c}{\# item \\ vertices} & 
			\tabincell{c}{\# user-item \\ edges} &
			\tabincell{c}{\# item-item \\ edges} &
			\tabincell{c}{\# attributes \\ of user} &
			\tabincell{c}{\# attributes \\ of item}
			\\
			\hline
			\textsl{Taobao-small} & 147,970,118 & 9,017,903 & 442,068,516 & 224,129,155 & 27 & 32 \\
			\textsl{Taobao-large} & 483,214,916 & 9,683,310 & 6,587,662,098 & 231,085,487 & 27 & 32\\
			\hline 
		\end{tabular}
	}
	\vspace{-1em}
\end{table}

\begin{figure}[h]
	\centering
	\includegraphics[width=2.7in]{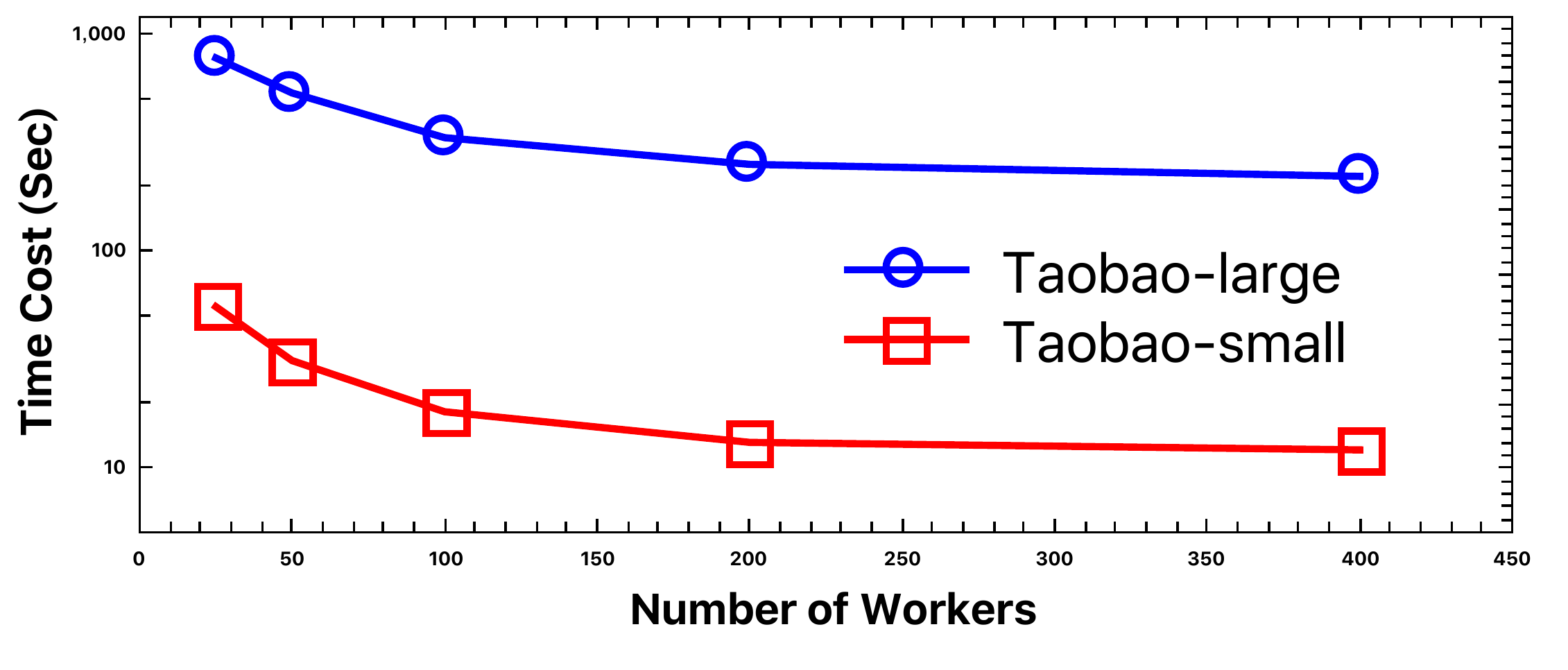}
	\vspace{-2em}
	\caption{Graph building time w.r.t.~number of workers. The graph could be built in several minutes on two datasets.
	}\label{fig:getime}
	\vspace{-1em}
\end{figure}

\smallskip
\noindent{\underline{\textbf{Graph Building.}}}
The performance of graph building plays a central role in a graph computation platform. AliGraph supports various kinds of raw data from different file systems, partitioned or not. Figure~\ref{fig:getime} presents the time cost of graph building w.r.t.~the number of workers on the two datasets.
We have the following two observations:
1) the graph building time explicitly decreases w.r.t.~the number of workers;
2) AliGraph can build large-scale graphs in minutes, e.g. 5 minutes for \textsl{Taobao-large}. This is much more efficient than most state-of-the-arts (e.g., PowerGraph~\cite{Gonzalez2012PowerGraph}) that usually takes several hours). 

\begin{figure}[h]
	\centering
	\includegraphics[width=2.8in]{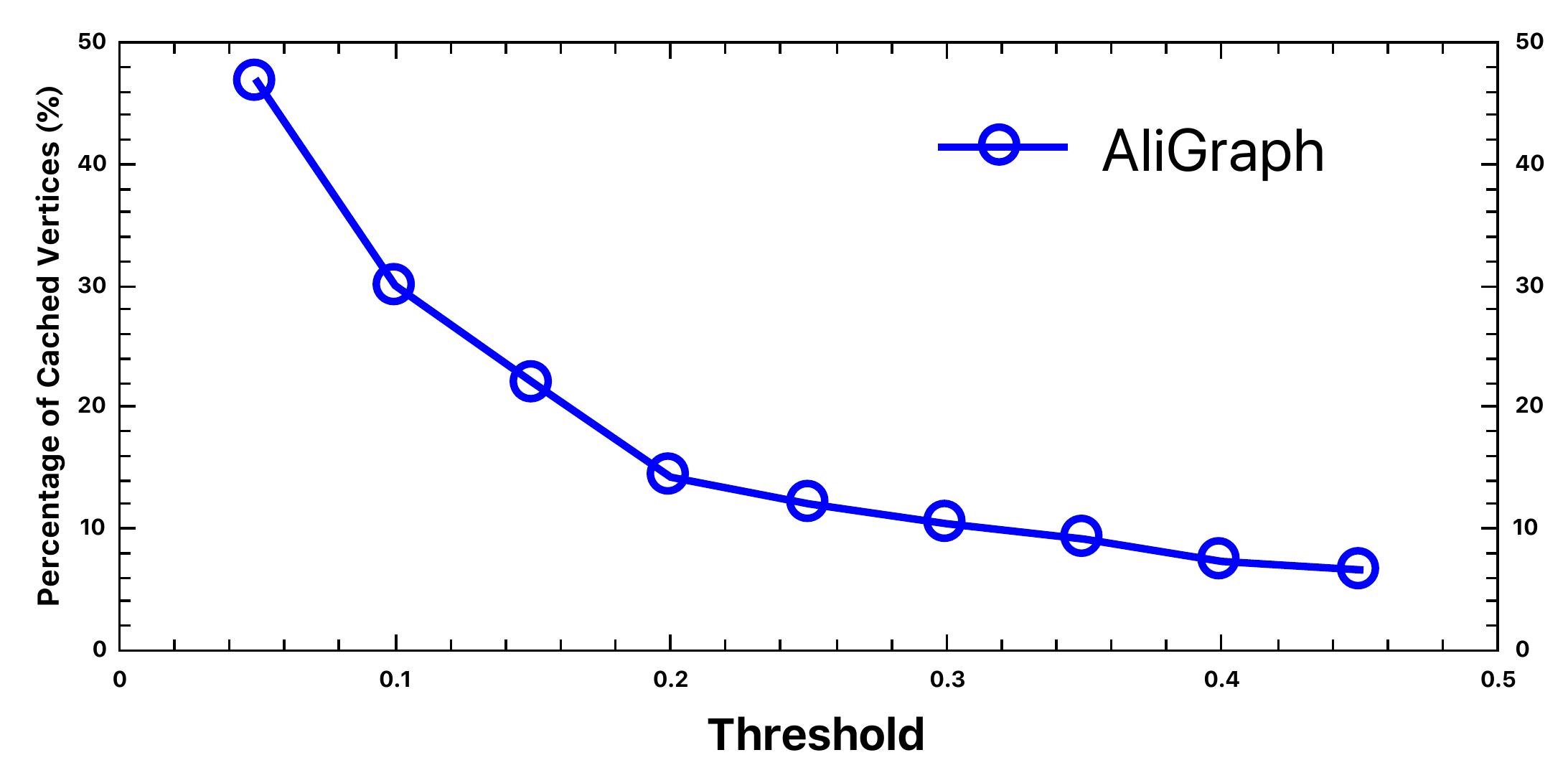}
	\vspace{-2em}
	\caption{Cache rate w.r.t.~threshold.
		Setting the threshold near $0.15$ makes the best trade-off.
	}\label{fig:cacheImp}
\end{figure}

\begin{figure}[h]
	\centering
	\includegraphics[width = 2.8in]{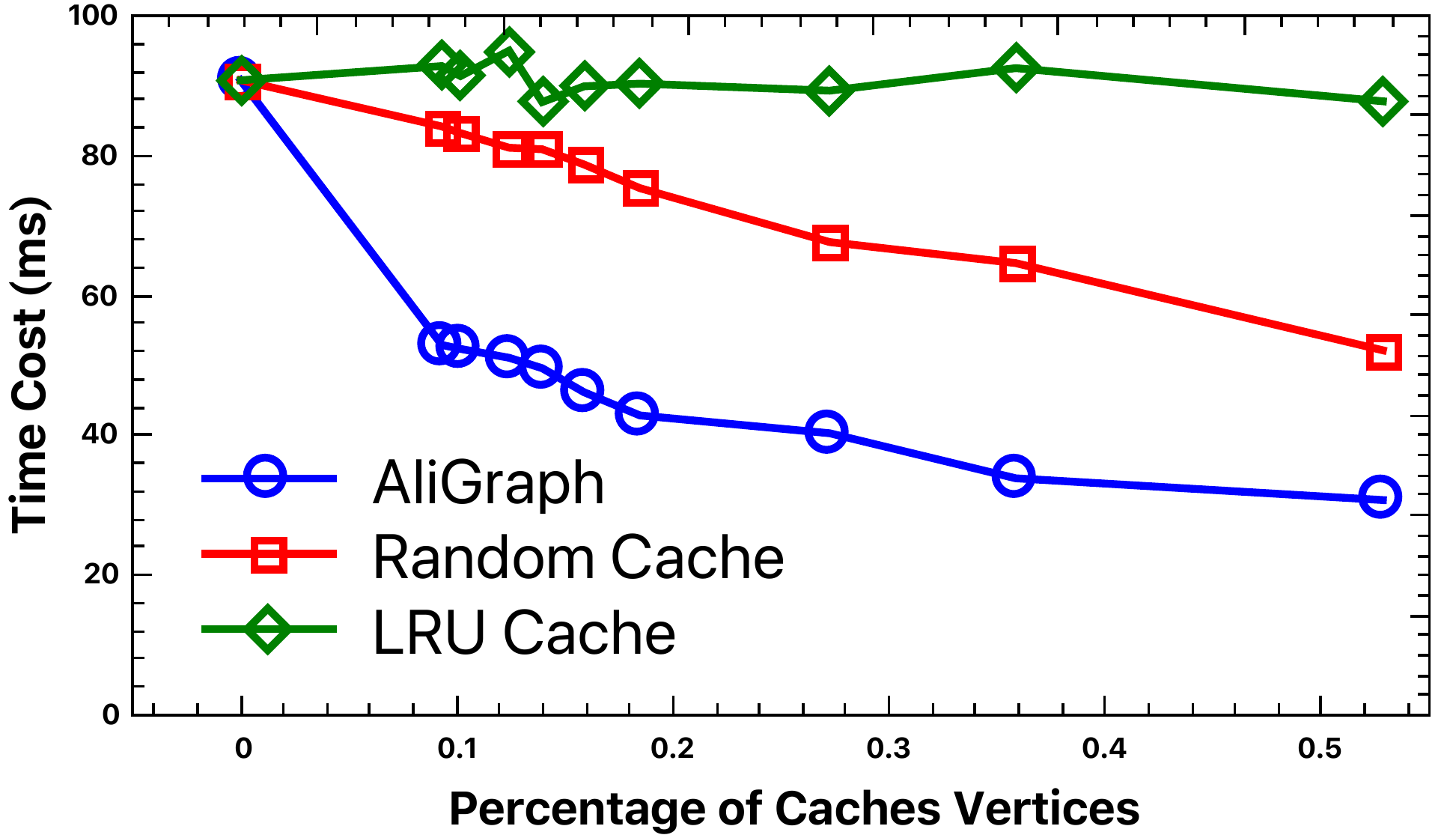}
	\vspace{-1em}
	\caption{Cost time w.r.t.~percentage of cached vertices. Our method saves much time than other cache strategies.}\label{fig:cacheTime}
	\vspace{-1.5em}
\end{figure}

\smallskip
\noindent{\underline{\textbf{Effects of Caching Neighbors.}}}
We examine the effects of caching $k$-hop neighbors of important vertices. 
In our caching algorithm, we need to set threshold for $Imp(v)$ as defined in Equation~\eqref{eq:Impkv} with the analysis of $D^{(k)}_i$ and $D^{(k)}_o$. In the experiments, we locally cache the $1$-hop (direct) neighbors of all vertices and vary the threshold controlling for caching the $2$-hop neighbors. We gradually increase the threshold from $0.05$ to $0.45$ to test its sensitivity and effectiveness. Figure~\ref{fig:cacheImp} illustrates the percentage of vertices being cached w.r.t.~the threshold. We observe that 
the percentage of cached vertices decreases w.r.t.~the threshold. When the threshold is smaller than $0.2$, it decreases drastically and becomes relatively stable after that. 
This is because the importance of vertices obey the power-law distribution as we prove in Theorem~\ref{Thm:ImpPowerlaw}. To make a good trade-off between the cache cost and the benefit, we set the threshold as $0.2$ based on Figure~\ref{fig:cacheTime} and only need to cache around $20\%$ of extra vertices. We also compare our importance-based caching strategy w.r.t.~two other strategies, namely the random strategy which caches the neighbors of a fraction of vertices selected at random and the LRU replacing strategy~\cite{Chrobak1998LRU}. Figure~\ref{fig:cacheTime} illustrates the cost time w.r.t.~the percentage of cached vertices. We find that our method saves about 40$\%$--50$\%$ time w.r.t.~the random strategy and about 50$\%$--60$\%$ time w.r.t.~the LRU strategy, respectively.
This is simply due to: 
1) the randomly selected vertices are less likely to be accessed; and 
2) the LRU strategy incurs additional cost since it frequently replaces cached vertices.
Whereas, our importance-based cached vertices are more likely to be accessed by others.

\smallskip
\noindent{\underline{\textbf{Effects of Sampling.}}} 
We test the effects of our optimized implementation on sampling with the batch size of $512$ and cache rate 20\%. Table~\ref{tab:system_sampling} shows the time cost of the three types of sampling methods. We find that:
1) Sampling methods are very efficient which finish between a few milliseconds to no more than 60ms;  
2) The sampling time grows slowly w.r.t.~the graph size. Although the storage size of \textsl{Taobao-large} is six times larger compared to \textsl{Taobao-small}, the sampling time on the two datasets is quite close.
These observations verify that our implementations of the sampling methods are efficient and scalable.

\begin{table}[t]
	\centering
	\caption{\label{tab:system_sampling} Effects of optimized \textsl{Sampling}. All sampling methods can finish in no more than $60$ms.
	}
	\resizebox{\columnwidth}{!}{
		\begin{tabular}{c|cc|ccc}
			\hline 
			\rowcolor{mygray}
			& \multicolumn{2}{c}{Setting} & 
			\multicolumn{3}{c}{Time (ms)} \\
			\cline{2-6}
			\rowcolor{mygray}
			\multirow{-2}{*}{Dataset} & $\#$ of workers & Cache Rate &
			\textsc{Traverse} & \textsc{Neighborhood}
			& \textsc{Negative}
			\\
			\hline
			\textsl{Taobao-small} & 25 & $18.46\%$ &
			\cellcolor{myemph} 2.59 & \cellcolor{myemph} 45.31 & 
			\cellcolor{myemph} 6.22 \\
			\textsl{Taobao-large} & 100 & $17.68\%$ &
			\cellcolor{myemph} 2.62 & 
			\cellcolor{myemph} 52.53 & 
			\cellcolor{myemph} 7.52 \\
			\hline 
		\end{tabular}
	}
	\vspace{-1em}
\end{table}

\smallskip
\noindent{\underline{\textbf{Effects of Operators.}}}
We further examine the effects of our implementation on the operators \textsc{Aggregate} and \textsc{Combine}. Table~\ref{tab:system_operator} shows the time cost of the two operators and the time costs can speed up by an order of magnitude with our proposed implementations.
This is simply because we apply the caching strategy to eliminate the redundant computation of intermediate embedding vectors. Once again, this verifies the superiority of our AliGraph platform.

\begin{table}[t]
	\centering
	\caption{\label{tab:system_operator} Effects of optimized \textsl{Operators} with an order of magnitude of time speed up. }
	\resizebox{\columnwidth}{!}{
		\begin{tabular}{c|cc|c}
			\hline 
			\rowcolor{mygray}
			Dataset & W/O Our Implementation (ms) &
			Our Implementation (ms) & Speedup Ratio
			\\
			\hline
			\textsl{Taobao-small} & 7.33 & \cellcolor{myemph} \textbf{0.57} & 12.9 \\
			\textsl{Taobao-large} & 17.21 &
			\cellcolor{myemph} \textbf{1.26} & 13.7   \\
			\hline 
		\end{tabular}
	}
	\vspace{-2em}
\end{table}

\subsection{Algorithm Evaluations}  

In this subsection, we evaluate the performance of our proposed GNNs compared to state-of-the-arts. 
We first describe the experimental settings including the datasets, competitors and evaluation metrics. Then, we examine the efficiency and effectiveness of each proposed GNN. 

\subsubsection{Experimental Settings}

\noindent{\underline{\textbf{Datasets.}}}
We employ two datasets in our experiments, including a public dataset from \textsl{Amazon} and \textsl{Taobao-small}. We choose \textsl{Taobao-small} due to the reason of the scalability of several competitors.

The statistics of the datasets are summarized in Table~\ref{tab:datasets}. Both of them are AHGs.
The public dataset \textsl{Amazon}
extracted from~\cite{mcauley2015image, he2016ups} is the product metadata under the electronics category of the Amazon company. 
In this graph, each vertex represents a product 
with its attributes and each edge connects two products co-viewed or co-bought by the same user. It has two types of vertices, namely user and item, and four types of edges between users and items, namely click, add-to-preference, add-to-cart and buy. 

\begin{table}[t]
	\centering
	\caption{\label{tab:datasets} Statistics of datasets used in experiments.}
	\resizebox{\columnwidth}{!}{
		\begin{tabular}{c|cccc}
			\hline 
			\rowcolor{mygray}
			Dataset & \tabincell{c}{\# of \\ vertices} & 
			\tabincell{c}{\# of \\ edges} &
			\tabincell{c}{\# of \\ vertex type} &
			\tabincell{c}{\# of \\ edge type}
			\\
			\hline
			\textsl{Amazon} & 10,166 & 148,865 & 1 & 2 \\
			\textsl{Taobao-small} & 156,988,021 & 666,197,671 & 2 & 4\\
			\hline 
		\end{tabular}
	}
	\vspace{-1em}
\end{table}

\smallskip
\noindent{\underline{\textbf{Algorithms.}}}
We implement all of our proposed algorithms in this paper. For comparison, we also implement some representative graph embedding algorithms in different categories as follows:

\noindent{\underline{\textit{C1: Homogeneous GE Methods.}}}
The compared methods include \textsf{DeepWalk}~\cite{perozzi2014deepwalk}, \textsf{LINE}~\cite{tang2015line}, and \textsf{Node2Vec}~\cite{grover2016node2vec}.
These methods can only be applied on plain graphs with purely structural information.

\noindent{\underline{\textit{C2: Attributed GE Methods.}}} The compared method includes \break \textsf{ANRL}~\cite{zhang2018anrl}, which can generate embeddings capturing both structural and attributed information.

\noindent{\underline{\textit{C3: Heterogeneous GE Methods.}}
	The compared methods include \textsf{Methpath2Vec}~\cite{dong2017metapath2vec}, \textsf{PMNE}~\cite{liu2017principled}, \textsf{MVE}~\cite{qu2017attention} and \textsf{MNE}~\cite{ijcai2018-428}.
	\textsf{Methpath2Vec} can only process graphs with multiple types of vertices while the other three methods can only process graphs with multiple types of edges. The \textsf{PMNE} involves three different
	kinds of approaches to extend the \textsf{Node2Vec} method, which are denoted as \textsf{PMNE-n}, \textsf{PMNE-r} and \textsf{PMNE-c}, respectively. }

\noindent{\underline{\textit{C4: GNN Based Methods.}}}
The comparison methods include \textsf{Structural2Vec}~\cite{Ribeiro2017struc2vec}, \textsf{GCN}~\cite{kipf2017semi}, \textsf{Fast-GCN}~\cite{chen2018fastgcn}, \textsf{AS-GCN}~\cite{huang2017accelerated}, \textsf{GraphSAGE}~
\cite{GraphSage:HamiltonYL17} and \textsf{HEP}~\cite{zheng2018heterogeneous}.

For fairness, all algorithms are implemented by applying the optimized operators on our system.
If a method cannot process attributes and/or multiple types of vertices, we simply ignore these information in the embedding. We generate the embedding for each subgraph with the same type of edges and concatenate them together to be the final result for homogeneous based GNN.
Notice that, in our examination, we do not compare each of our proposed GNN algorithms w.r.t.~all competitors. This is because each algorithm is designed with different focus. We will detail the competitors of each GNN algorithm in reporting its experimental results.

\smallskip
\noindent{\underline{\textbf{Metrics.}}}
We evaluate both the efficiency and effectiveness of the proposed methods. The efficiency can be simply measured by the execution time of the algorithm. To measure effectiveness, following previous work~\cite{Cai2017A, cui2018survey,Hamilton2017Representation}, we apply the algorithm on the widely adopted link prediction task, which plays important roles in real-world scenarios such as recommendation.
We randomly extract a portion of the data as the training data and reserve the remaining part as test data. To measure the quality of the results, four commonly used metrics are applied, namely the area under ROC curve (\texttt{ROC-AUC}), the PR curve (\texttt{PR-AUC}), the \texttt{$F_1$-score} and the hit recall rate (\texttt{HR Rate}). Notably, each metric is averaged among different types of edges. 

\smallskip
\noindent{\underline{\textbf{Parameters.}}}
We set $d$, the dimension of embedding vectors, to be $200$ for all algorithms.

\subsubsection{Experimental Results}

We report the detailed experimental results of each proposed GNN algorithm here.

\begin{figure}[t]
	\centering
	\includegraphics[width = 0.47\columnwidth]{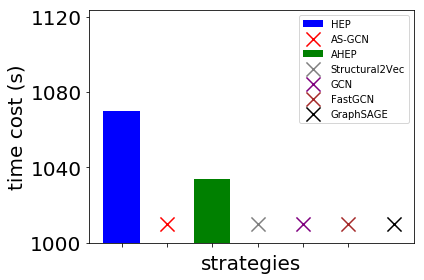}
	\includegraphics[width = 0.47\columnwidth]{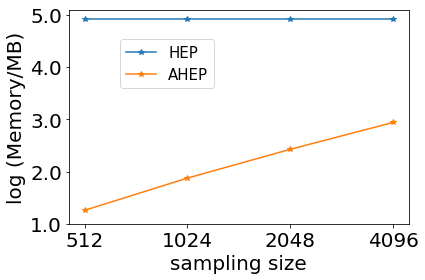}
	\vspace{-2mm}
	\caption{Average memory cost and running time of per batch. $\times$ indicates the algorithm can not terminate in reasonable time. \textsf{AHEP} is 2--3 faster than \textsf{HEP} and uses much less memory on \textsl{Taobao-small}.
	}\label{fig:hep-memtime}
	\vspace{-1em}
\end{figure}

\smallskip
\noindent{\underline{\textbf{\textsf{AHEP} Algorithm.}}}
The goal of the \textsf{AHEP} algorithm is to fast obtain the embedding result while does not sacrifice too much accuracy. In Table~\ref{tab:exp-effc-ahep}, we show the comparison results on result quality of \textsf{AHEP} w.r.t.~its competitors on the \textsl{Taobao-small} dataset.  In Figure~\ref{fig:hep-memtime}, we illustrate time and space cost of different algorithms. Obviously, we have the following observations:
1) On the large \textsl{Taobao-small} dataset, \textsf{HEP} and \textsf{AHEP} are the only two algorithms that can produce results in reasonable time and space limits.
However, \textsf{AHEP} is about $2$--$3$ faster than \textsf{HEP} and uses much less memory than \textsf{HEP}.
2) In terms of the result quality, the \texttt{ROC-AUC} and \texttt{$F_1$-score} of \textsf{AHEP} is comparable to \textsf{HEP}. 
These verify that \textsf{AHEP} can produce similar results of \textsf{HEP} by using much less time and space.

\begin{table}[t]
	\centering
	\small
	\caption{\label{tab:exp-effc-ahep} 
		Effectiveness comparison of \textsf{AHEP} w.r.t.~its competitors. \textsf{AHEP} is close to \textsf{HEP} on \textsl{Taobao-small}.}
	\smallskip
	{
		\begin{tabular}{c|ccc|ccc}
			\hline
			\rowcolor{mygray}
			Method & \texttt{ROC-AUC($\%$)} & \texttt{$F_1$-score($\%$)} \\
			\hline
			\textsf{Structural2Vec} & N.A. & N.A. \\
			\textsf{GCN} & N.A. & N.A. \\
			\textsf{FastGCN} & N.A. & N.A. \\
			\textsf{GraphSAGE} & N.A. & N.A. \\
			\textsf{AS-GCN} & O.O.M & O.O.M \\
			\textsf{HEP} & 77.77 & 57.93 \\
			\textsf{AHEP} &75.51 & 50.97 \\
			\hline
		\end{tabular}
	}
	\small
	\\
	{
		``N.A.'' indicates the algorithm can not terminate in reasonable time. ``O.O.M.'' indicates that the algorithm terminates due to out of memory.
	}
	\vspace{-2em}
\end{table}

\smallskip
\noindent{\underline{\textbf{\textsf{GATNE} Algorithm.}}}
The goal of \textsf{GATNE} is designed to process graphs with heterogeneous and attributed information on both vertices and edges. We show the comparison results of the \textsf{GATNE} algorithm w.r.t.~its competitors in Table~\ref{tab:exp-effc-gatne}. Obviously we find that \textsf{GATNE} outperforms all existing methods in terms of all metrics. For example, on the \textsl{Taobao-small} dataset, \textsf{GATNE} improves the \texttt{ROC-AUC}, \texttt{PR-AUC} and \texttt{$F_1$-score} by $4.6\%$, $1.84\%$ and $5.08\%$, respectively.
This is simply due to that \textsf{GATNE} simultaneously captures both the heterogenous information of vertices and edges and the attributes information. Meanwhile, we find that training time of \textsf{GATNE} decreases almost linearly w.r.t.~the number of workers. The \textsf{GATNE} model converges in less than 2 hours with 150 distributed workers. The verifies the high efficiency and scalability of the \textsf{GATNE} method.

\begin{table*}[t]
	\centering
	\caption{\label{tab:exp-effc-gatne} 
		Effectiveness comparison of \textsf{GATNE} w.r.t.~its competitors. \textsf{GATNE} outperforms all competitors in terms of all metrics on both \textsl{Amazon} and \textsl{Taobao-small}.
		\textsf{GATNE} lifts the \texttt{$F_1$-score} by $16.43\%$ on the \textsl{Amazon} dataset.
	}
	\smallskip
	\resizebox{0.7\textwidth}{!}
	{
		\begin{tabular}{c|ccc|ccc}
			\hline
			\rowcolor{mygray}
			& \multicolumn{3}{c}{\textsl{Amazon}} & \multicolumn{3}{c}{\textsl{Taobao-small}} \\
			\rowcolor{mygray}
			\multirow{-2}{*}{Method} & \texttt{ROC-AUC($\%$)} & \texttt{PR-AUC($\%$)} & \texttt{$F_1$-score($\%$)} & \texttt{ROC-AUC($\%$)} & \texttt{PR-AUC($\%$)} & \texttt{$F_1$-score($\%$)}\\
			\hline
			\textsf{DeepWalk} & 94.20 & 94.03 & 87.38 & 65.58 & 78.13 & 70.14\\
			\textsf{Node2Vec} & 94.47 & 94.30 & 87.88 &
			N.A. & N.A. & N.A.\\
			\textsf{LINE} & 81.45 & 74.97 & 76.35 &
			N.A. & N.A. & N.A.\\ \hline
			\textsf{ANRL} & 95.41 & 94.19 & 89.60 &
			N.A. & N.A. & N.A.\\ \hline
			\textsf{Metapath2Vec} & 94.15 & 94.01 & 87.48 & N.A. & N.A. & N.A. \\
			\textsf{PMNE-n} & 95.59 & 95.48 & 89.37 &
			N.A. & N.A. & N.A.\\
			\textsf{PMNE-r} & 88.38 & 88.56 & 79.67 &
			N.A. & N.A. & N.A.\\
			\textsf{PMNE-c} & 93.55 & 93.46 & 86.42 &
			N.A. & N.A. & N.A.\\
			\textsf{MVE} & 92.98 & 93.05 & 87.80 & 
			66.32 & 80.12 & 72.14  \\
			\textsf{MNE} & 91.62 & 92.46 & 84.44 & 
			79.60 & 93.01 & 84.86 \\
			\rowcolor{myemph}
			\textsf{GATNE} & \textbf{96.25} & \textbf{94.77} & \textbf{91.36} &
			\textbf{84.20} & \textbf{95.04} & \textbf{89.94} \\\hline
		\end{tabular}
	}
	\vspace{-2em}
\end{table*}

\smallskip
\noindent{\underline{\textbf{\textsf{Mixture GNN}.}}}
We compare our \textsf{Mixture GNN} method w.r.t. \textsf{DAE}\break~\cite{Vincent2008Extracting} and \textsf{$\beta^{*}$-VAE}~\cite{Liang2018Variational} methods. The hit recall rate of applying the embedding results into the recommendation task is shown in Table~\ref{tab:exp-effc-MixtureGNN}. Notice that, by applying our model, the hit recall rates have been improved by around $2\%$. Similarly, this improvement also makes significant contributions in a large network. 

\smallskip
\noindent{\underline{\textbf{\textsf{Hierarchical GNN}.}}}
We compare our \textsf{Hierarchical GNN} method w.r.t.~\textsf{GraphSAGE}. The results is shown in Table~\ref{tab:exp-effc-HieGNN}. The \texttt{$F_1$-score} is significantly improved by around $7.5\%$. This indicates that our \textsf{Hierarchical GNN} can generate more promising embedding results.

\smallskip
\noindent{\underline{\textbf{\textsf{Evolving GNN}.}}}
We compare our \textsf{Evolving GNN} method w.r.t. other methods on the multi-class link prediction task. The competitors include the representative algorithms \textsf{DeepWalk}, \textsf{DANE}, \textsf{DNE}, \textsf{TNE} and \textsf{GraphSAGE}. These competitor algorithms can not handle dynamic graphs, thus we run the algorithm on each snapshot of the dynamic graphs and report the average performance over all timestamps. The comparison results on the \textsl{Taobao-small} dataset are shown in Table~\ref{tab:exp-effc-evolveGNN}. We easily find that, \textsf{Evolving GNN} outperforms all other methods in terms of all metrics. For example, with burst change, \textsf{Evolving GNN} improves the micro and macro \texttt{$F_1$-score} by $4.2\%$ and $3.6\%$. This is simply because our proposed method can better capture the dynamic changes of real-world networks, thus can produce more promising results.

\begin{table}[t]
	\centering
	\small
	\caption{\label{tab:exp-effc-MixtureGNN} 
		Effectiveness comparison of \textsf{Mixture GNN} w.r.t.~its competitors. \textsf{Mixture GNN} improves the hit recall rate by around $2\%$ on \textsl{Taobao-small}.}
	\smallskip
	{
		\begin{tabular}{c|cc}
			\hline
			\rowcolor{mygray}
			Method & \texttt{HR Rate$@20$} & \texttt{HR Rate$@50$} \\
			\hline
			\textsf{DAE} & 0.12622 & 0.21619 \\
			\textsf{$\beta^{*}$-VAE} & 0.11767 & 0.19997 \\
			\rowcolor{myemph}
			\textsf{Mixture GNN} & \textbf{0.14317} & \textbf{0.23680} \\
			\hline
		\end{tabular}
	}
	\vspace{-1em}
\end{table}

\begin{table}[t]
	\centering
	\small
	\caption{\label{tab:exp-effc-HieGNN} 
		Effectiveness comparison of \textsf{Hierarchical GNN} w.r.t.~its competitors. \textsf{Hierarchical GNN} improves the hit recall rate by $7.5\%$  on \textsl{Taobao-small}.}
	\smallskip
	\resizebox{\columnwidth}{!}
	{
		\begin{tabular}{c|ccc}
			\hline
			\rowcolor{mygray}
			Method & \texttt{ROC-AUC$(\%)$}  & \texttt{PR-AUC$(\%)$} & \texttt{$F_1$-score$(\%)$}   \\
			\hline
			\textsf{GraphSAGE} & 82.89 & 44.45 & 45.76 \\
			\rowcolor{myemph}
			\textsf{Hierarchical GNN} & \textbf{87.34} & \textbf{54.87} & \textbf{53.20} \\
			\hline
		\end{tabular}
	}
	\vspace{-2em}
\end{table}

\begin{table*}[t]
	\centering
	\small
	\caption{\label{tab:exp-effc-evolveGNN} 
		Effectiveness comparison of \textsf{Evolving GNN} w.r.t.~its competitors. \textsf{Evolving GNN} improves the \texttt{$F_1$-score} by about $4\%$  on \textsl{Taobao-small}.}
	\smallskip
	\resizebox{0.7\textwidth}{!}
	{
		\begin{tabular}{c|cc|cc}
			\hline
			\rowcolor{mygray}
			& \multicolumn{2}{c}{Normal Evolution} & \multicolumn{2}{c}{burst Change} \\
			\rowcolor{mygray}
			\multirow{-2}{*}{Method} & Micro \texttt{$F_1$-score($\%$)} & Macro \texttt{$F_1$-score($\%$)} & Micro \texttt{$F_1$-score($\%$)} & Macro \texttt{$F_1$-score($\%$)}\\
			\hline
			\textsf{DeepWalk} & N.A. & N.A. & N.A. & N.A. \\
			\textsf{DANE} & N.A. & N.A. & N.A. & N.A. \\
			\textsf{TNE} & 79.9 & 71.9 & 69.1 & 67.2 \\
			\textsf{GraphSAGE} & 71.4 & 70.4 & 60.7 & 60.5 \\
			\rowcolor{myemph}
			\textsf{Evolving GNN} & \textbf{81.4} & \textbf{77.7} & \textbf{73.3} & \textbf{70.8} \\
			\hline
		\end{tabular}
	}
	\vspace{-2em}
\end{table*}

\smallskip
\noindent{\underline{\textbf{\textsf{Bayesian GNN}.}}}
The goal of this model is to combine Bayesian method with the traditional GNN model. We use   \textsf{GraphSAGE} as the baseline and compare the results with and without incorporating the proposed Bayesian model. We present the hit recall rate of the recommendation result in Table~\ref{tab:exp-effc-BysGNN}. Notice that, we considerthe granularity of both item brands and categories. Obviously, when applying our Bayesian model, the hit recall rates have been increased by $1\%$ to $3\%$ respectively. Notice that, this improvement can bring significant benefits on our network containing 9 million items.

\begin{table*}[t]
	\centering
	\small
	\caption{\label{tab:exp-effc-BysGNN} 
		Effectiveness comparison of \textsf{Bayesian GNN} w.r.t.~its competitors. \textsf{Bayasian GNN} improves the hit recall rate by $1\%$--$3\%$ on \textsl{Taobao-small}.}
	\smallskip
	\begin{tabular}{c|c|cc|cc}
		\hline
		\rowcolor{mygray}
		& & \multicolumn{2}{c}{Click} & \multicolumn{2}{c}{Buy} \\
		\rowcolor{mygray}
		\multirow{-2}{*}{Granularity} & 
		\multirow{-2}{*}{\texttt{HR Rate}} &
		\textsf{GraphSAGE} &
		\textsf{GraphSAGE + Bayesian} &
		\textsf{GraphSAGE} &
		\textsf{GraphSAGE + Bayesian} \\ \hline
		& 10 & 15.97 & \cellcolor{myemph} \textbf{16.14} & 24.87 & \cellcolor{myemph} \textbf{25.10} \\
		Brand & 30 & 16.65 & \cellcolor{myemph} \textbf{17.12} & 25.70 & \cellcolor{myemph} \textbf{26.57} \\
		& 50 & 17.26 & \cellcolor{myemph} \textbf{17.90} & 26.39 & \cellcolor{myemph} \textbf{27.33} \\ \hline
		& 10 & \textbf{27.46} & \cellcolor{myemph} 27.49 & 27.85 & \cellcolor{myemph} \textbf{27.91} \\
		Category & 30 & 28.43 & \cellcolor{myemph} \textbf{29.99} & 28.50 & \cellcolor{myemph} \textbf{29.45} \\
		& 50 & 29.58 & \cellcolor{myemph} \textbf{32.88} & 26.26 & \cellcolor{myemph} \textbf{31.47} \\ \hline
	\end{tabular}
	\vspace{-1em}
\end{table*}

\section{Related Work}\label{sec:relatedwork}
In this section, we briefly review the state-of-the-arts on GE and GNN methods. Based on the four challenges summarized in Section~1, we categorize existing methods  as follows.

\smallskip
\noindent{\underline{\textbf{Homogeneous.}}}
\textsf{DeepWalk}~\cite{perozzi2014deepwalk} first generates a corpus on graphs by random walk and then trains a skip-gram model on the corpus. LINE~\cite{tang2015line} learns node presentations by preserving both first-order and second-order proximities. NetMF~\cite{qiu2018network} is a unified matrix factorization framework for theoretically understanding and improving DeepWalk and LINE.  Node2Vec \cite{grover2016node2vec} adds two parameters to control the random walk process while SDNE~\cite{wang2016structural} proposes a structure-preserving embedding method. GCN~\cite{kipf2017semi} incorporates neighbors' feature representations using convolutional operations.
GraphSAGE~\cite{GraphSage:HamiltonYL17} provides an inductive approach to combine structural information with node features.  

\smallskip
\noindent{\underline{\textbf{Heterogeneous.}}}
For graph with multiple types of vertices and/or edges, PMNE~\cite{liu2017principled} proposes three methods to project a multiplex network into a continuous vector space. MVE~\cite{qu2017attention} embeds networks with multiple views in a single collaborated embedding using the attention mechanism. MNE~\cite{ijcai2018-428} uses one common embedding and several additional embeddings of each edge-type for each node, which are jointly learned by a unified network embedding model. Mvn2Vec~\cite{oneshotrelational:Xiong} explores the embedding results by simultaneously modeling preservation and collaboration. HNE~\cite{chang2015heterogeneous} jointly considers the contents and topological structures to be 
unified vector representations. PTE~\cite{tang2015pte} constructs large-scale heterogeneous text network from  labeled information, which is then embedded into a low-dimensional space. Metapath2Vec~\cite{dong2017metapath2vec} and HERec~\cite{shi2018heterogeneous} formalize meta-path based random-walks to construct the heterogeneous neighborhood of a node and then leverage skip-gram models to perform node embeddings. 

\smallskip
\noindent{\underline{\textbf{Attributed.}}}
Attributed network embedding aims to seek for low-dimensional vector representations to preserve both  topological and attribute information.
TADW~\cite{yang2015network} incorporates text features of vertices into network representation learning by matrix
factorization. 
LANE~\cite{huang2017label} smoothly incorporates label information into the attributed network embedding while preserving their correlations.
AANE~\cite{huang2017accelerated} enables joint learning process to be done in a distributed manner for accelerated attributed network embedding.
SNE~\cite{liao2018attributed} proposes a generic framework for embedding social networks by capturing both the structural proximity and attribute proximity. 
DANE~\cite{gao2018deep} can capture the high nonlinearity and preserve various proximities in both topological structure and node attributes. ANRL~\cite{zhang2018anrl} uses a neighbor enhancement autoencoder to model the node attribute information and a skip-gram model to capture the network structure.

\smallskip
\noindent{\underline{\textbf{Dynamic.}}}
Actually, some static methods \cite{perozzi2014deepwalk,tang2015line} can also handle dynamic network by updating the new vertices based on static embedding. Considering the new vertices' influence on the original networks, \cite{dne} extends the skip-gram methods to update the original vertices' embedding. \cite{triad} focuses on capturing the triadic structure properties for learning network embedding. Considering both the network structure and node attributes, \cite{attributed} focuses on updating the top eigenvectors and eigenvalues for the streaming network.

\section{Conclusions and Future Work}\label{sec:future}
We summarize four challenges from the current practical graph data problems, namely \emph{large-scale}, \emph{heterogeneous}, \emph{attributed} and \emph{dynamic}. Based on these challenges, we design and implement a platform, \textit{AliGraph}, which provides both system and algorithms to tackle more practical problems. In the future, we will focus on but not limited to the following directions: 1) GNN for edge-level and subgraph-level embeddings; 2) More execution optimizations, such as co-location of computation variables in GNN with graph data to reduce the cross network traffic, introduction of new gradient optimization  to leverage the trait of GNN to speed up the distributed training without accuracy loss, and better assignment of the workers  in multi-GPU architectures; 3) Early-stop mechanism, which can help to terminate training tasks earlier when no promising results can generate; 4) Auto-ML, which can help to select the optimal method from a variety of GNNs.

\bibliographystyle{abbrv}
\bibliography{reference}


\setcounter{theorem}{0}

\vspace{-1em}
\section*{Appendix}

\textbf{Proof of Theorem~1}
Let $D_{i}^{(k)}$ and $D_{o}^{(k)}$ be two random variables representing the number of $k$-hop in and out-neighbors of a randomly chosen vertex from the graph, respectively. We derive the probability distribution of $D_{i}^{(k)}$ and $D_{o}^{(k)}$ for each $k \geq 1$ by induction. 

1. Following previous work~\cite{Tanimoto2009Power}, when $k = 1$, the in-degree $D_{i}^{(1)}$ and out-degree $D_{o}^{(t)}$ both obey the power-law distribution. Specifically, let $\gamma^{(1)}_{o}$ and $\gamma^{(1)}_{i}$ denote the exponent, we have $\Pr(D_{i}^{(1)} = q) \propto q^{-\gamma^{(1)}_{i}}$ and $\Pr(D_{o}^{(1)} = q) \propto q^{-\gamma^{(1)}_{o}}$.

2. Then, we consider the probability distribution of $D_{i}^{(k)}$ and $D_{o}^{(k)}$ where $k \geq 2$. 
Assume that $D_{i}^{(k-1)}$ and $D_{o}^{(k-1)}$ obey the power-law distribution with exponent $\gamma_{i}^{(k-1)}$ and $\gamma_{o}^{(k-1)}$, respectively. Let $u$ be a randomly chosen vertex from the graph. If we randomly chose a $(k-1)$-hop in-neighbor $s$ of $u$ and a one-hop out-neighbor $v$ of $u$, $v$ is obviously a $k$-hop out-neighbor of $s$. Since $s$ is chosen randomly, we have
\vspace{-1em}
\begin{equation*}
\begin{split}
~ & \Pr(D_{o}^{(k)} = q) = \sum_{j=1}^{n} \Pr(D_{o}^{(1)} = j) \cdot \Pr(D_{i}^{(k)} = \frac{q}{j})\\
& \propto \sum_{j=1}^{n} {j}^{-\gamma^{(1)}_{o}}
{(\frac{q}{j})}^{-\gamma^{(k-1)}_{i}} \propto q^{-\gamma^{(k-1)}_{i}} 
\sum_{j=1}^{n} 
{j}^{\gamma^{(k-1)}_{i} - \gamma^{(1)}_{o}}.
\end{split}
\end{equation*}

Since $n$, $\gamma^{(t-1)}_{i}$ and $\gamma^{(1)}_{o}$ are all fixed, the last term is a constant value. As a result, we have  $\Pr(D_{o}^{(k)} = q) \propto q^{-\gamma^{(k)}_{o}}$,
where $\gamma^{(k)}_{o}$ is a term determined by $\gamma^{(k-1)}_{i}$ and the last term. 
In similar, we also have $\Pr(D_{i}^{(k)} = q) \propto q^{-\gamma^{(k)}_{i}}$. This indicates both $D_{o}^{(k)}$ and $D_{i}^{(k)}$ obey the power-law distribution.

By summarizing 1 and 2, we find that both $D_{i}^{(k)}$ and $D_{o}^{(k)}$ obey the power-law distribution for each $k \geq 1$.

\textbf{Proof of Theorem~2}
Let $Imp^{(k)} = D_{i}^{(k)} / D_{o}^{(k)}$ be a random variable denoting the importance of a randomly chosen vertex. We have
\vspace{-1em}
\begin{equation*}
 \begin{split}
~ & \Pr({Imp}^{(k)} = \ell) = \sum_{j = 1}^{n}
    \Pr(D_{o}^{(k)} = j) \Pr(D_{i}^{(k)} = \ell j) \\
& = \sum_{j = 1}^{n} j^{-\gamma_{o}^{(k)}} 
{(\ell j)}^{-\gamma_{i}^{(k)}} = {\ell}^{-\gamma_{i}^{(k)}}
\sum_{j = 1}^{n} j^{-(\gamma_{i}^{(k)} + \gamma_{o}^{(k)})}.
\end{split}   
\end{equation*}
Since the last term is also a constant value, 
we find that $Imp^{(k)}$ also obeys the power-law distribution. This analysis result indicates that the importance $Imp^{(k)}$ of most vertices is very small. Thus, we only need to cache a small number of vertices with large importance values. Intuitively, for any vertex whose importance is large, it has a large number of in-neighbors and a small number of out-neighbors. 

\end{document}